\documentclass[a4paper,11pt]{article}
\pdfoutput=1
\usepackage{jcappub}

\usepackage{lscape}
\usepackage{fancyhdr}
\usepackage[font={footnotesize}]{caption}
\usepackage[english]{babel}
\usepackage{datetime} 
\usepackage{titlesec}
\usepackage{enumitem}
\usepackage[section]{placeins}
\usepackage{nicefrac}
\usepackage{graphicx}
\usepackage{subfig}

\newcommand{\deriv}[2]{\frac{\mathrm{d}#1}{\mathrm{d}#2}}
\newcommand{\sderiv}[2]{\frac{\mathrm{d}^2#1}{\mathrm{d}#2^2}}	
\newcommand{\pderiv}[2]{\frac{\partial #1}{\partial #2}}	
\newcommand{\psderiv}[3]{\frac{\partial ^2#1}{\partial #2\partial #3}}
\newcommand{\psderiva}[2]{\frac{\partial ^2#1}{\partial #2^2}}

\newdateformat{dateUKenglish}{\THEDAY~\monthname[\THEMONTH] \THEYEAR}
\numberwithin{equation}{section}
\title{Weak lensing deflection of three-point correlation functions}
\author[1]{Susan Pyne,}
\author[1]{Benjamin Joachimi,}
\author[1,2]{Hiranya V. Peiris}
\affiliation[1]{Department of Physics and Astronomy,
University College London,
Gower Street,
London WC1E 6BT,
UK}
\affiliation[2]{Oskar Klein Centre for Cosmoparticle Physics,
Stockholm University,
AlbaNova,
Stockholm SE-106 91,
Sweden}
\emailAdd{susan.pyne.15@ucl.ac.uk}
\emailAdd{b.joachimi@ucl.ac.uk}
\emailAdd{h.peiris@ucl.ac.uk}
\abstract{    Weak gravitational lensing alters the apparent separations between observed sources, potentially affecting clustering statistics. We derive a general expression for the lensing deflection which is valid for any three-point statistic, and investigate its effect on the three-point clustering correlation function.
We find that deflection of the clustering correlation function is greatest at around $z=2$. It is most prominent in regions where the correlation function varies rapidly, in particular at the baryon acoustic oscillation scale where it smooths out the peaks and troughs, reducing the peak-to-trough difference  by about 0.1 percent at $z=1$ and around 2.3 percent at $z=10$.   The modification due to lensing deflection is typically at the per cent level of the expected errors in a Euclid-like survey and therefore undetectable.     }
\notoc

\begin{document}
\maketitle
\flushbottom
\newpage

\section{Introduction}\label{intro}
On scales larger than galaxies the structure of the Universe depends only on properties of primordial inhomogeneities and their subsequent evolution under gravity. The resulting matter distribution is commonly quantified through clustering statistics such as the two-point correlation function (2PCF) or, in Fourier space, the power spectrum. The primordial density distribution is determined by inflation and is expected to be nearly Gaussian and therefore fully described by these two-point statistics. However, later gravitational collapse of overdense regions causes coupling between different Fourier modes and makes the matter distribution non-Gaussian. As a result information is transferred into higher-order statistics, which are complementary to two-point statistics.  For example, the three-point correlation function (3PCF) or bispectrum has  been used to investigate primordial non-Gaussianity \cite{RN61, RN8}, to estimate galaxy bias (the differential clustering of galaxies compared with dark matter) \cite{RN26}, and to constrain cosmological parameters, in particular $\Omega_\mathrm{m}$ and $\sigma_8$ \cite{RN27}.

Since the 1970s two- and three-point correlation functions have been measured with increasing accuracy in galaxy surveys such as the Two-degree-Field Galaxy Redshift Survey\footnote{http://www.2dfgrs.net/} and the Sloan Digital Sky Survey (SDSS).\footnote{http://www.sdss.org/} A recent aim has been the detection of baryon acoustic oscillations (BAOs) in clustering statistics. The position of BAO peaks acts as a standard ruler which can provide information about distances and hence about the expansion of the Universe.    The first evidence for BAO peaks in the 3PCF was reported in 2009 \cite{RN43}. Subsequently  evidence at the $2.8\sigma$ level was reported using data from the SDSS Baryon Oscillation Spectroscopic Survey (BOSS) \cite{RN34}, and  more recently the first high confidence  detection ($4.5\sigma$) was reported, also using SDSS data \cite{RN63}.      Future surveys such as the Dark Energy Spectroscopic Instrument (DESI) \cite{RN41},  Euclid\footnote{http://sci.esa.int/euclid/} \cite{RN88} and the Large Synoptic Survey Telescope (LSST) \cite{RN98}  will improve upon these measurements, for example by halving the uncertainties in distance measurements at the BAO scale \cite{RN41}. Consequently there will be a need for increasingly accurate modelling of the 2PCF and 3PCF.

Theoretical expressions for the matter 2PCF and 3PCF up to second order in the density contrast can be derived using Newtonian  perturbation theory \cite{RN5} and are sufficient for comparison with current surveys. However other contributions to the correlation functions may also be important for future surveys. For example a detailed analysis of the effect of weak lensing magnification on the galaxy 3PCF \cite{RN12} concluded that this effect was potentially detectable in galaxy and quasar samples at $z=3$ in future \lq{}ideal\rq{} surveys.       More recent work has emphasised that future wide and deep surveys will require  the inclusion of additional relativistic terms in the power spectrum and bispectrum of galaxy number counts    \cite{RN42,RN94,RN95, RN29, RN96,RN64}.     These terms arise because at higher redshifts the quantities which we observe, such as positions, volumes and densities, differ from their true source values, due to the propagation of light through the inhomogeneous matter distribution along the line of sight.  Some effects  which are first order in the density and volume in perturbation theory are detectable in current surveys, for example redshift-space distortions, weak lensing and the integrated Sachs-Wolfe effect.  Many second order effects have also been estimated but have not been shown to be detectable in current or planned surveys \cite{RN29}.

One second order relativistic effect due to lensing is the alteration of the apparent distance between two or more patches of the sky so that sources are not observed at their true positions \cite{RN42}. For galaxy samples this is a much smaller effect than the more commonly studied weak lensing magnification and shear.  However it is potentially larger than other relativistic contributions and its implications for two-point and higher-order statistics could be relevant for future surveys.  

 Early studies of deflection due to weak lensing mainly considered the effect on cosmic microwave background (CMB) anisotropies \cite {RN66,RN65}. Since then the effect of lensing deflection on the power spectrum of CMB temperature anisotropies has  been explored in detail in harmonic space \cite{RN52,RN7,RN82,RN6}, demonstrating that lensing is a non-trivial contaminant of CMB  temperature and polarisation  observations but also introduces valuable additional information.  In a wider context Ref. \cite{RN1} estimated the impact of deflection on the matter 2PCF, assuming two sources at the same redshift.     They concluded that the effect is small  except where the correlation function is rapidly changing, for example near the BAO feature in galaxy surveys \cite{RN2}, where they estimated the effect to be around the percent level.   Here lensing deflection tends to wash out the details of the peaks and troughs. Importantly, this deflection does not affect the position of the BAO peaks although it sets a limit on the accuracy with which the amplitudes of the peaks can be measured \cite{RN2}. Thus it has no implications for the use of the BAO scale as a standard ruler.  More recently Ref.  \cite{RN75} repeated this analysis in Fourier space and confirmed the size of the smoothing effect on the matter power spectrum at the BAO scale, also showing that at $z=4$ this effect has approximately the same magnitude as smoothing due to non-linear structure.  
Even if the deflection effect is not important for two-point statistics, it could potentially make a measurable, and interesting, contribution to three-point  statistics. In harmonic space  expressions have been developed for the lensed CMB bispectrum \cite{RN82, RN6}. In this work we instead extend the general real-space analysis in Ref. \cite{RN1} to the 3PCF,  now with three  sources  at the same redshift, and consider whether the effect could be detected in forthcoming surveys.  

This paper is organised as follows: Section 2 derives general expressions for the lensed 3PCF and the associated lensing deflection; Section 3 applies our new derivations to the 3PCF of the matter density field; Section 4 discusses the observability of the deflection effect; Section 5 contains our conclusions. Detailed derivations are given in appendices.  Throughout we assume a flat $\Lambda$CDM universe.

\section{Lensed three-point correlation function }\label{3PCF}
 Suppose that a physical observable $A(\mathbf{x}_a)$ is  observed at  position $\mathbf{x}_a$. The true position is not as observed because photons are deflected as they travel to the observer.
Thus the (lensed) quantity $\tilde{A}(\mathbf{x}_a)$ which is observed at $\mathbf{x}_a$  is actually at a  different position $\mathbf{x}_a+\boldsymbol{\lambda}_a$, where $\boldsymbol{\lambda}_a$ is a deflection vector:
\begin{equation}
	\tilde{A}(\mathbf{x}_a)= A(\mathbf{x}_a+\boldsymbol{\lambda}_a)\  .\label{eq:A}
\end{equation}
If we measure a correlation function of the observable we necessarily measure the correlation between lensed variables, which is not the same as the true correlation function.
 Taking the 2PCF as an example,
\begin{align}
\langle \tilde{A}(\mathbf{x}_a)\tilde{B}(\mathbf{x}_a)\rangle&= \langle A(\mathbf{x}_a+\boldsymbol{\lambda}_a)B(\mathbf{x}_b+\boldsymbol{\lambda}_b)\rangle\\
&\ne \langle A(\mathbf{x}_a)B(\mathbf{x}_a) \rangle\ .
\end{align}

Our approach to deriving an expression for the lensed 3PCF is motivated by results showing that in the two-point case the lensed 2PCF can be expressed as the sum of the unlensed 2PCF and  a lensing deflection term,  $\langle AB\rangle_2$ \cite{RN1}.   Importantly, the transverse deflection is much greater than the deflection along the line of sight so in this work we consider only  transverse displacements. 

 Assuming that the comoving distances to $A$ and $B$  are approximately the same  and that the deflection is small so a perturbative approach can be used, the lensed 2PCF is given by
\begin {align}
	\langle \tilde A \tilde B \rangle&= \langle AB \rangle+ \langle AB \rangle_2\\
    &\approx \langle AB \rangle +\frac{1}{r}\left(T-\frac{D}{2}\right)\deriv {\langle AB\rangle}{r}+ \left(T+\frac{D}{2}\right)\sderiv{\langle AB\rangle}{r}\ ,\label{eq:AB}
 \end{align}
 where $r$ is the distance between $A$ and $B$. The functions $T$ and $D/2$ are respectively the trace and off-diagonal traceless part of a distortion tensor with components $Z_{ij}$, where $i$  and $j$ denote two orthogonal directions in the plane of the sky:

\begin{align}
Z =\begin{pmatrix}
	T+\frac{D}{2} & 0\\
	0 & T-\frac{D}{2}  \label{eq:Zij}
\end{pmatrix} .
\end{align}

 The components of the deflection vector can be expressed in terms of integrals of the gravitational potential over the line of sight which arise as solutions of the geodesic equations (see for example Ref. \cite{RN6} for  a derivation in conformal Newtonian gauge). In a flat $\Lambda$CDM universe the two orthogonal transverse components of $\boldsymbol{\lambda}_a$ (denoted by the index $l=1,2$) are given by \cite{RN3}
\begin {align}
	\lambda_{a,\perp}^l &= \frac{2}{c^2}\int_0^{\chi_a}\mathrm{d}\chi(\chi_a-\chi)\nabla^l \Phi(\chi)\ , 
\label{eq:BS1}
\end{align}
where $\chi_a$ is the comoving distance to the source and $\Phi(\chi)$ is the gravitational potential.

\vspace{1cm}
\begin{figure}[t]
\centering
\vspace*{-0.5cm}
\fbox{\includegraphics[scale=0.4]{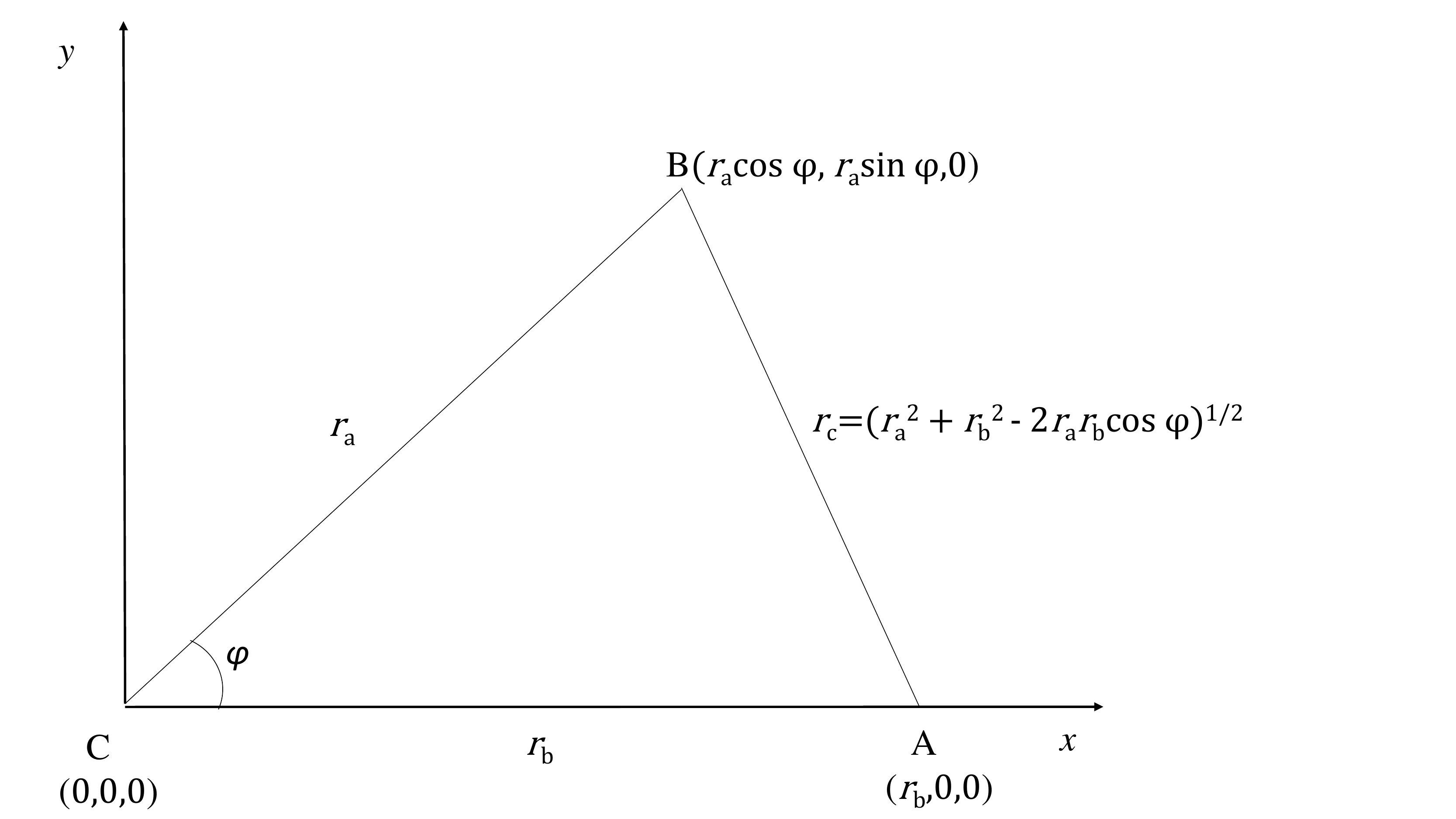}}
\caption{Choice of coordinates for the three points. Point A is at $(r_{b1}, r_{b2},0)= (r_b,0,0)$, point B is at  $(r_{a1}, r_{a2},0) =(r_a\cos\varphi, r_a\sin\varphi,0)$ and point C is at the origin.}\label{fig:coords3} 
\end{figure}

To obtain an expression for the lensed 3PCF, firstly  we assume that all three sources are at the same redshift. This is justified by the 2PCF finding that the dominant lensing effect is in the transverse direction, which means that triangles oriented closer to the line of sight will be less affected by lensing. 
Because of isotropy it is possible to choose coordinates with the $x$-axis along $\mathbf{x}_a\--\mathbf{x}_c$, the $y$-axis along the line of sight to $\mathbf{x}_c$,  and all three points in the $x\--y$ plane. The triangle formed from the three points  can then be defined in terms of two sides, $\mathbf{r}_a=(r_{a1}, r_{a2},0)$ and  $\mathbf{r}_b=(r_{b1}, r_{b2},0)$,  and the angle $\varphi$ between them, as shown in Figure \ref{fig:coords3}. Thus $\mathbf{r}_a$ is the observed distance between $\mathbf{x}_b$ and $\mathbf{x}_c$, and  $\mathbf{r}_b $ is the observed distance between $\mathbf{x}_a$ and $\mathbf{x}_c$. 

 As explained in Appendix \ref{L3PCF},  using Eq. (\ref{eq:A}) we can write the lensed 3PCF  in the same way as for the 2PCF as the sum of the unlensed correlation function and a lensing deflection term:
\begin{align}
	\langle\tilde{A}\tilde{B}\tilde{C}\rangle&=\langle ABC\rangle +\langle ABC\rangle_2 \label{eq:ABC1}\ .
\end{align}
Following Ref. \cite{RN1}, we define three distortion tensors, $Z_{ac}$, $Z_{ab}$  and $Z_{bc}$, each quantifying the deflection along one side of the triangle, with elements
\begin{align}
	Z_{\alpha\beta}^{ij}\equiv \frac{(\langle\lambda_\alpha^i\lambda_\alpha^j\rangle+\langle\lambda_\beta^i\lambda_\beta^j\rangle)}{2}-\langle\lambda_\alpha^i\lambda_\beta^j\rangle\ ,\label{eq:Z}
\end{align}
where $\alpha\beta$ is $ab$, $bc$  or $ca$.
 We assume that the lensing deflection is small so that terms above second order can be neglected, and that the observables are not correlated with the lensing deflection field.  Then by Taylor-expanding the expression  for the lensed correlation function, Eq. (\ref{eq:ABC1}), we can write the lensing deflection in terms of the deflection  tensors.  Appendix \ref{L3PCF} gives fuller details. The final result is
\begin{align}
	\langle ABC\rangle_2&= \psderiv{\langle ABC\rangle}{r_{bi}}{r_{bj}}Z_{ac}^{ij}\notag\\
	&\quad+ \psderiv{\langle ABC\rangle}{r_{ai}}{r_{aj}}Z_{bc}^{ij}\notag\\
	&\quad+\psderiv{\langle ABC\rangle}{r_{ai}}{r_{bj}}[Z_{ac}^{ij}+Z_{bc}^{ij}-Z_{ab}^{ij}]\ ,\label{eq:ABC2}
\end{align}
 where the indices $i,j$ take the values  1 and 2, denoting the two transverse directions,  and repeated indices are summed over. In the chosen coordinates $Z_{ab}^{12} =Z_{ab}^{21}=0$. Thus we only need to consider the $i=j=1$ and $i=j=2$ terms and Eq. (\ref{eq:ABC2}) simplifies to 
\begin{align}
	\langle ABC\rangle_2&= \psderiva{\langle ABC\rangle}{r_{b1}}Z_{ac}^{11}+\psderiva{\langle ABC\rangle}{r_{a1}}Z_{bc}^{11}+\psderiva{\langle ABC\rangle}{r_{a2}}Z_{bc}^{22}\notag\\
	&\quad\quad+\psderiv{\langle ABC\rangle}{r_{a1}}{r_{b1}}[Z_{ac}^{11}+Z_{bc}^{11}-Z_{ab}^{11}]\ .
\end{align}
\\
As shown in Appendix \ref{L3PCF}, this can be expressed in terms of derivatives of $r_a$, $r_b$ and $\varphi$ only:
\begin{align}
	\langle ABC\rangle_2&=  \psderiva{\langle ABC \rangle}{r_a}\left[Z_{bc}^{11}\cos^2\varphi + Z_{bc}^{22}\sin^2 \varphi \right]\notag\\
	&\quad+\psderiva{\langle ABC \rangle}{r_b}Z_{ac}^{11}\notag\\
	&\quad+\psderiva{\langle ABC \rangle}{\varphi}\left[\frac{Z_{bc}^{11}\sin^2\varphi+Z_{bc}^{22}\cos^2\varphi}{r_a^2}\right]\notag\\
	&\quad+\psderiv{\langle ABC \rangle}{r_a}{r_b}\cos\varphi[Z_{ac}^{11}+Z_{bc}^{11}-Z_{ab}^{11}]\notag \\
	&\quad-\psderiv{\langle ABC \rangle}{r_a}{\varphi}\frac{2\sin\varphi\cos\varphi}{r_a}\left[Z_{bc}^{11}-Z_{bc}^{22}\right]\notag\\		
	&\quad-\psderiv{\langle ABC \rangle}{r_b}{\varphi}\frac{\sin\varphi}{r_a}[Z_{ac}^{11}+Z_{bc}^{11}-Z_{ab}^{11}]\notag \\		
	&\quad+\pderiv{\langle ABC \rangle}{r_a}\left[\frac{ Z_{bc}^{11}\sin^2 \varphi+ Z_{bc}^{22} \cos^2 \varphi}{r_a}\right]\notag\\	
	&\quad+\pderiv{\langle ABC \rangle}{\varphi}\frac{2\sin\varphi\cos\varphi}{r_a^2}\left[Z_{bc}^{11}-Z_{bc}^{22}\right].\label{eq:ABC2*}
\end{align} 

 To determine $Z_{ac}^{ij}$, $Z_{ab}^{ij}$  and $Z_{bc}^{ij}$ we follow the arguments of Ref. \cite{RN1} for the 2PCF.  Using Eq. (\ref{eq:Zij}) each tensor $Z_{\alpha\beta}$  can be written in terms of its trace $T_{\alpha\beta}^\prime$ plus an off-diagonal traceless part $D_{\alpha\beta}^\prime/2$. The functions $T^\prime$ and $D^\prime$ are related to the functions $T$ and $D$ in Eq. (\ref{eq:AB}).

From Eqs. (A17) and (A18) of Ref.  \cite{RN1} we have
\begin{align}
	\langle\lambda_\alpha^i \lambda_\beta^j\rangle
 	&=T_{\alpha\beta}\delta_{ij} -\frac{D_{\alpha\beta}}{r^2}\left[r_{i}r_{j}-\frac{r^2}{2}\delta_{ij}\right] \ ,\label{eq:zetaab}
\end{align}
where $\delta_{ij}$ is the Kronecker delta.  

Now consider $Z_{ac}^{11}$ as an example. From Eq. (\ref{eq:Z}) this is defined as 
\begin{align}
	Z_{ac}^{11} &= \frac{\langle\lambda_a^1\lambda_a^1\rangle+\langle\lambda_c^1\lambda_c^1\rangle}{2}-\langle\lambda_a^1\lambda_c^1\rangle\ .\\
          \notag
\end{align}
Using Eq. (\ref{eq:zetaab})  and noting that $D_{aa}=D_{cc}=0$ \cite{RN1}, this can be written as
\begin{align}
	Z_{ac}^{11} 
	&= \frac{1}{2}\left(T_{aa}+T_{cc} \right)- \left(T_{ac}-\frac{D_{ac}^\prime}{r_b^2}\left(r_{b1}^2 -\frac{r_b^2}{2}\right)\right)\\
          &=T_{ac}^\prime + \frac{D_{ac}^\prime}{2}\ .
\end{align}
\\
The $i=j=1$ and $i=j=2$ elements of the other tensors can be derived in a similar way. They are:
\begin{align}
	Z_{bc}^{11}	&=T_{bc}^\prime+\frac{D_{bc}^\prime}{r_a^2}\left[r_{a1}^2-\frac{r_a^2}{2}\right]\notag\\
          &= T_{bc}^\prime+D_{bc}^\prime\left[\cos^2\varphi-\frac{1}{2}\right]\ ,\\
	Z_{bc}^{22}	&=T_{bc}^\prime+\frac{D_{bc}^\prime}{r_a^2}\left[r_{a2}^2-\frac{r_a^2}{2}\right]\notag\\
          &= T_{bc}^\prime+D_{bc}^\prime\left[\sin^2\varphi-\frac{1}{2}\right]\ ,\\
	Z_{ab}^{11}&= T_{ab}^\prime+D_{ab}^\prime\frac{[2r_{c1}^2-r_c^2]}{2r_c^2}\notag\\
	&= T_{ab}^\prime+D_{ab}^\prime\frac{[2(r_b-r_a\cos\varphi)^2-(r_a^2+r_b^2-2r_ar_b\cos \varphi)]}{2(r_a^2+r_b^2-2r_ar_b\cos\varphi)}\ .
\end{align}
\\
 The functions $T_{\alpha\beta}^\prime$ and $D_{\alpha\beta}^\prime$ are derived in Ref. \cite{RN1} using Eq. (\ref{eq:zetaab}) together with the Limber approximation, which is valid since the integration kernels are broad (as with cosmic shear):
\begin{align}
	T_{\alpha\beta}^\prime(\chi_0,r)	&=\frac{1}{c^2}\int_0^{\chi_0}\mathrm{d}\chi(\chi_0-\chi)^2\int_0^\infty\frac{k^3 \mathrm{d}k}{\pi} P_\Phi(k,\chi)\big[1-J_0(kr\chi/\chi_0)\big]\ , \\
		D_{\alpha\beta}^\prime(\chi_0,r)&= \frac{2}{c^2}\int_0^{\chi_0}\mathrm{d}\chi(\chi_0-\chi)^2
		\int_0^\infty\frac{k^3\mathrm{d}k}{\pi}P_\Phi(k,\chi)J_2(kr\chi/\chi_0) \ ,\label{eq:Dprime}
\end{align}
where $\chi_0$ is the comoving distance to the plane containing the three points, $P_\Phi(k,\chi)$ is the power spectrum of the gravitational potential, and $\alpha\beta = ac$, $bc$ or $ba$.   $P_\Phi(k,\chi)$  is related to the matter power spectrum, $P_\delta(k,\chi)$,  by
\begin{align}
	P_\Phi(\chi,k)	&=\frac{9}{4}\frac{H_0^4\Omega_\mathrm{m}^2}{k^4a^2} P_\delta(\chi,k)\ ,
\end{align}
where $a$ is the scale factor.

Equation (\ref{eq:ABC2*}) is a completely general result which makes no assumptions about the nature of the observables or the Gaussianity of either the observed field or the lensing potential. In contrast similar analysis for the matter power spectrum \cite{RN75}  and the CMB \cite{RN6} makes the simplifying assumption that the lensing potential is Gaussian.  Ref. \cite{RN75} estimate that the relative error in the power spectrum due to the assumption of Gaussianity is around $10^{-3}$ for scales and redshifts of interest. The error increases with redshift and is larger at small scales where perturbation theory may no longer be valid.

Equation (\ref{eq:ABC2*}) shows that the deflection effect depends on derivatives of the unlensed 3PCF. This means it will be most significant if the correlation function is rapidly varying, for example near the BAO feature in the matter 3PCF.      The dependence on derivatives of the 3PCF also means that the result is independent of galaxy bias so long as we can assume that bias is linear.   Linear bias would not be a valid assumption for precision modelling of the BAO since the BAO scale is well within the weakly nonlinear regime.   However it is a justifiable assumption for our more broadbrush estimates. The assumption of linear bias could in fact be valid for much smaller scales: for angular galaxy clustering it is possible to construct a linear bias model which is valid down to scales well into the nonlinear regime, possibly as small as $1\  h^{-1}$ Mpc \cite{RN70}.     
\section{Results} \label{results}
\subsection{Preliminaries}
  In this section we apply the results of Section \ref{3PCF} to the three-point correlation function, $\zeta(\mathbf{p},\mathbf{q},\mathbf{s})$,  of the matter density contrast $\delta(\mathbf{x})$, defined as
\begin{align}
\zeta(\mathbf{p},\mathbf{q},\mathbf{s})= \langle \delta(\mathbf{x}) \delta(\mathbf{x}+\mathbf{p})\delta(\mathbf{x}+\mathbf{q}) \rangle_{\mathbf{x} }\  ,
\end{align}  
where the subscript $\mathbf{x}$ indicates the average over all spatial positions, and  $\mathbf{p}$, $\mathbf{q}$ and $\mathbf{s}$ form the sides of a triangle.
We compute the unlensed  3PCF using second-order Eulerian perturbation theory \cite{RN4}. Details are given in Appendix \ref{U3PCF}.  All results are based on a flat $\Lambda$CDM cosmology with $\Omega_\mathrm{m}=0.3$, $\Omega_\mathrm{b}=0.05$, $h =0.7$ and $\sigma_8=0.9$, and use the nonlinear matter power spectrum from Ref. \cite{RN51}.

In Section \ref{comp} we compare the unlensed 3PCF, $\zeta$, with the lensed 3PCF, $\tilde{\zeta}$, defined by Eq. (\ref{eq:ABC1}), across a range of scales and at different redshifts. We present the lensing deflection $|\zeta-\tilde{\zeta} |$, and also the relative deflection $|\zeta-\tilde{\zeta}|/\zeta$. 
 To exemplify the properties of the 3PCF and the lensing deflection we give results only for triangles with two equal sides ($r_1=r_2$) and focus on two illustrative triangle shapes: equilateral triangles ($r_1=r_2=r_3$) and a specific \lq{}squeezed\rq{} shape with two equal sides with angle $\varphi =5$ degrees between them. 
Section \ref{redshift} discusses the redshift dependence of the lensing deflection, and in Section \ref{BAO} we present results at the BAO scale $80 \leq r \leq 120 \  h^{-1}$ Mpc.

\subsection {Comparison between lensed and unlensed 3PCF}\label{comp}
\begin{figure}[t]
\centering
\text{             Equilateral}
\hspace{6cm}
\text{Squeezed}
\subfloat{\includegraphics[width=0.5\textwidth]{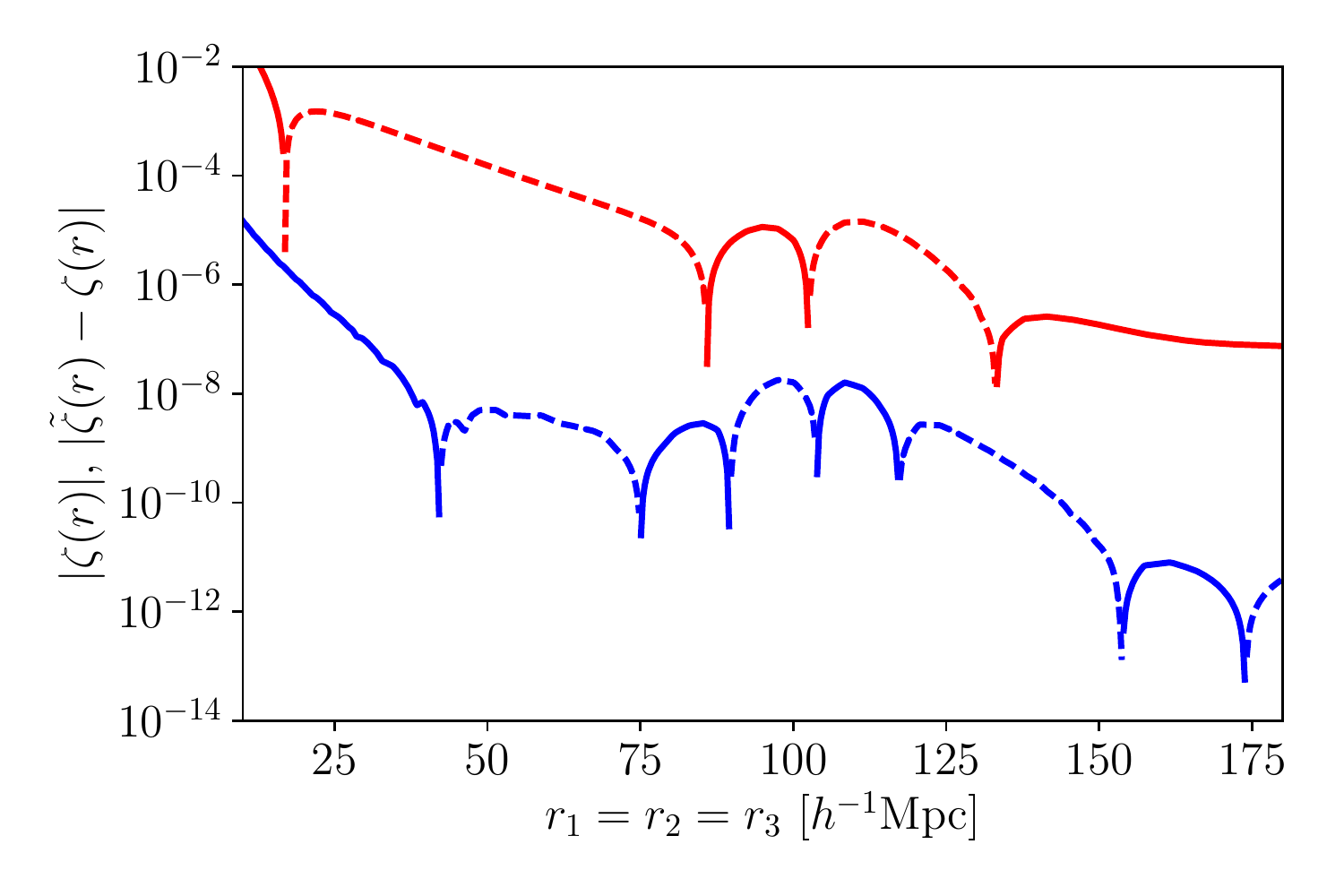}}
\hfill
\subfloat{\includegraphics[width=0.5\textwidth]{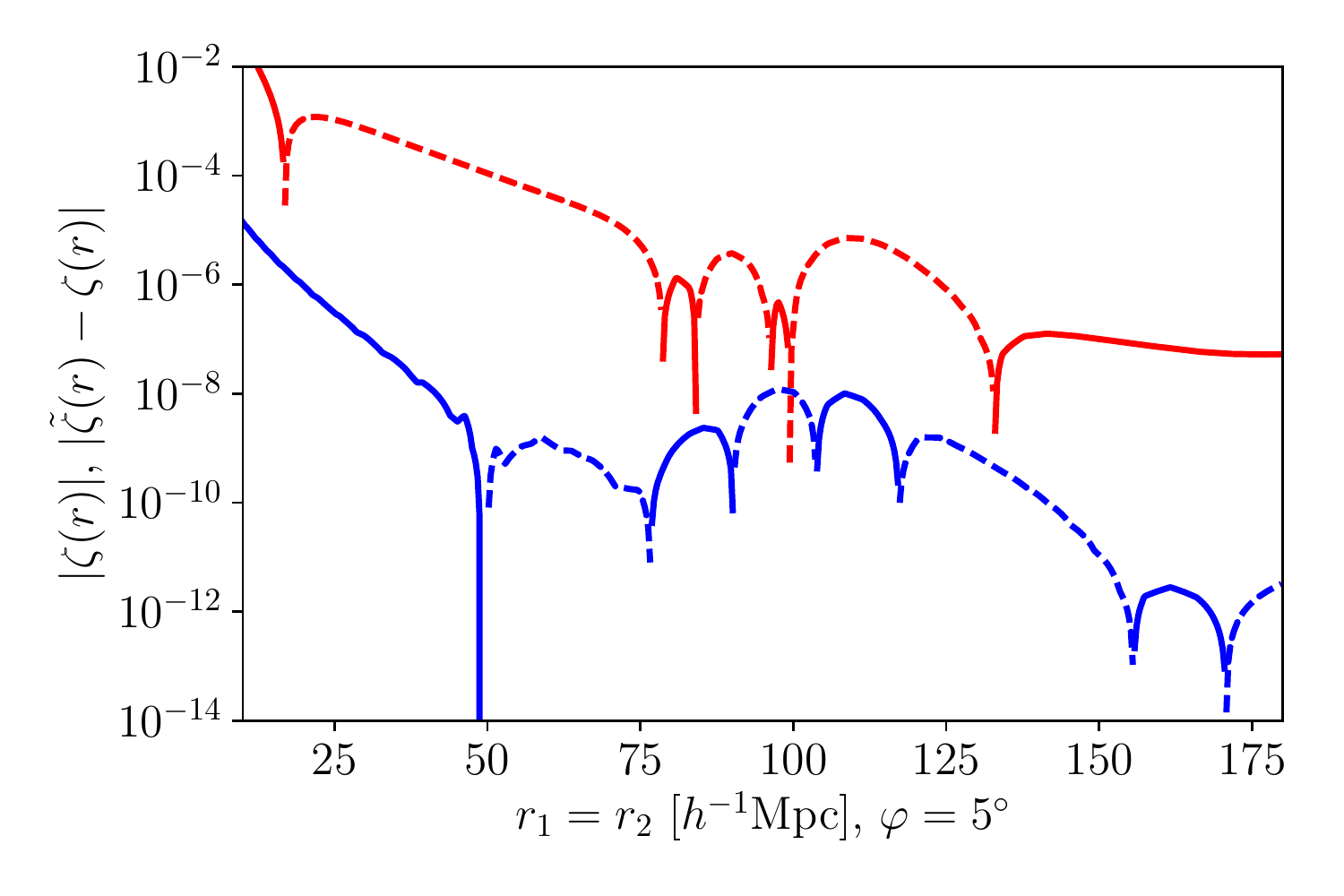}}
\caption {Unlensed three-point correlation function (red) and lensing deflection (blue) at  $z=1.0$. 
\textit{ Left}: Equilateral triangles. \textit{Right}: Squeezed triangles with $r_1 = r_2$ and $\varphi=5$ degrees. Dashed lines indicate negative values.} \label{fig:z1}
\end{figure}

Figure \ref{fig:z1} compares the magnitudes of the lensed 3PCF and the lensing deflection at $z=1$ for equilateral and squeezed triangles.  At this redshift, typical of current and planned galaxy surveys, the absolute value of the lensing deflection is around $10^{-8}$.  In general, the lensing effect is larger at small scales because the photon paths are more highly correlated.  However the relative contribution is strongest near the BAO feature where the partial derivatives in Eq. (\ref{eq:ABC2*}) are large.

\begin{figure}[t]
\centering
\includegraphics[width=0.5\textwidth]{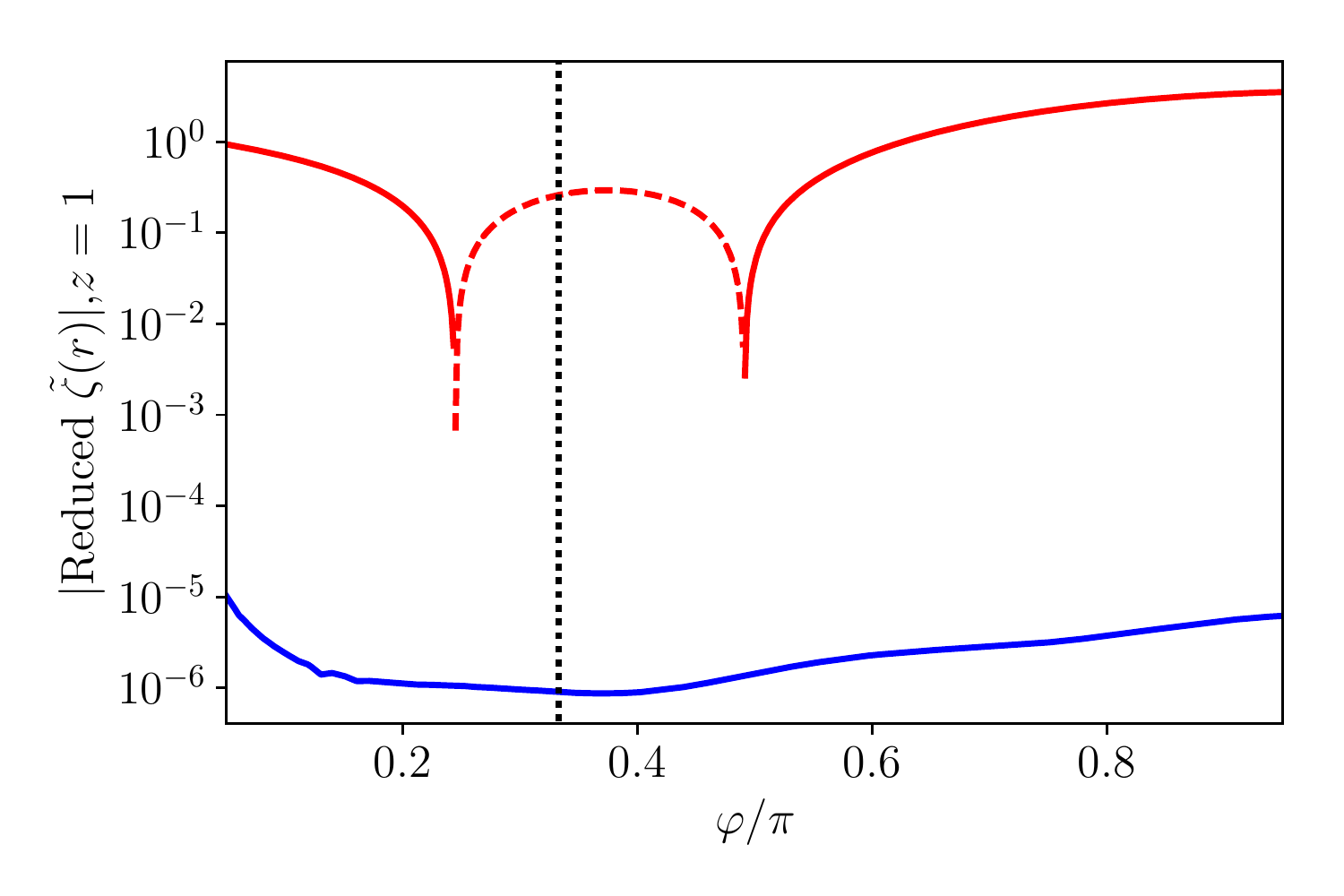}
   \caption{Reduced lensed three-point correlation function (red) and lensing deflection (blue)  at $z=1$ for triangles with $r_1=r_2=20\  h^{-1}$ Mpc as a function of angle $\varphi$ between these sides. The vertical line marks the position of equilateral triangles. Dashed lines indicate negative values.} \label{fig:red}
\end{figure}

To explore the shape, rather than size, of the 3PCF it is convenient to define the reduced (or normalised) 3PCF, $Q$ \cite{RN82}, which is essentially independent of redshift. It is defined as 
\begin{align}
 	Q=\frac{\zeta(\mathbf{p},\mathbf{q},\mathbf{s})}{\xi(\mathbf{p},\mathbf{q}) +\xi(\mathbf{q},\mathbf{s})+\xi(\mathbf{s},\mathbf{p}) }\ , \label{eq:Q}
\end{align}
where $\xi$ is the two-point correlation function. The reduced lensed 3PCF can be defined similarly, with the lensed 3PCF in the numerator and  lensed 2PCFs in the denominator.  
Figure \ref{fig:red} shows how the reduced lensed 3PCF at $z=1$ varies with angle $\varphi$ between two equal sides $r_1=r_2= 20\  h^{-1}$ Mpc.  The scale chosen is illustrative of a small scale away from the BAO feature; similar results apply at other scales. The 3PCF attains a minimum for approximately equilateral triangles and increases as the length of the third side decreases  (squeezed triangles) or increases (flattened triangles).  Lensing has least effect on the 3PCF of equilateral triangles because the correlation function  is relatively smooth and lacking in detail; lensing has most effect when the 3PCF is rapidly changing.
\subsection{Lensing deflection as a function of redshift}\label{redshift}
\begin{figure}[t]
\centering
\text{             Equilateral}
\hspace{6cm}
\text{Squeezed}
\subfloat{\includegraphics[width=0.48\textwidth]{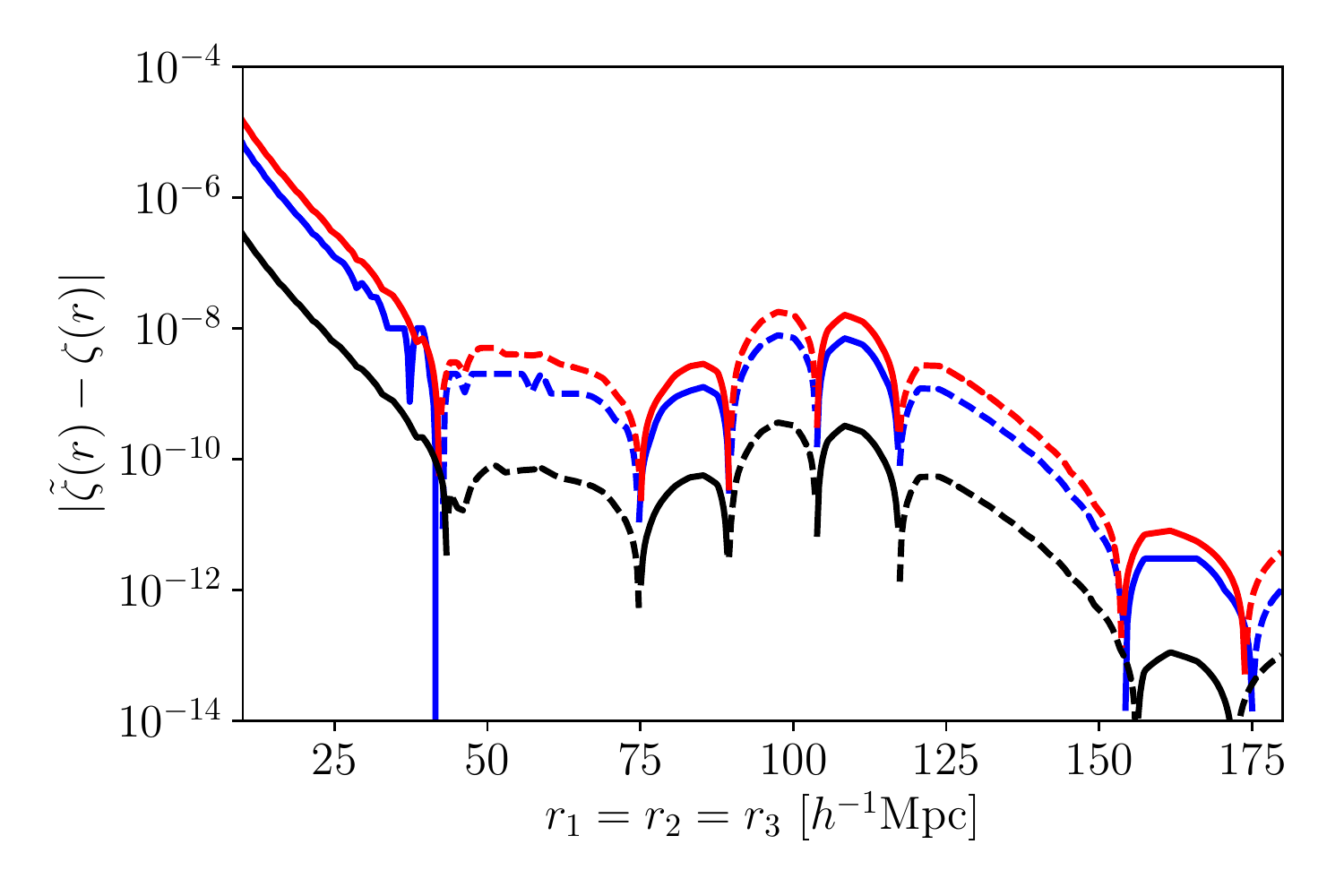}}
  \hfill
\subfloat{\includegraphics[width=0.48\textwidth]{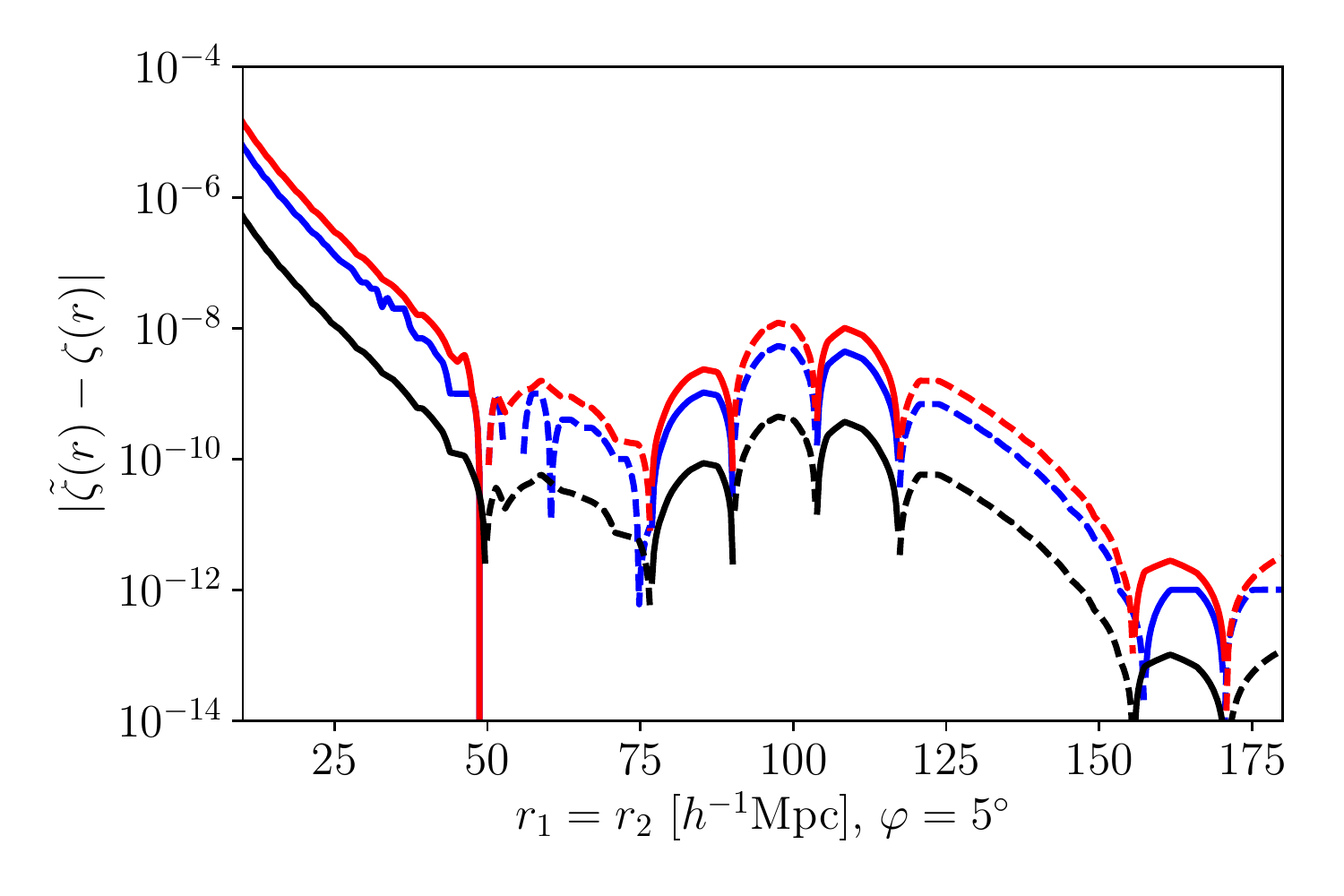}}
\vfill
\subfloat{\includegraphics[width=0.48\textwidth]{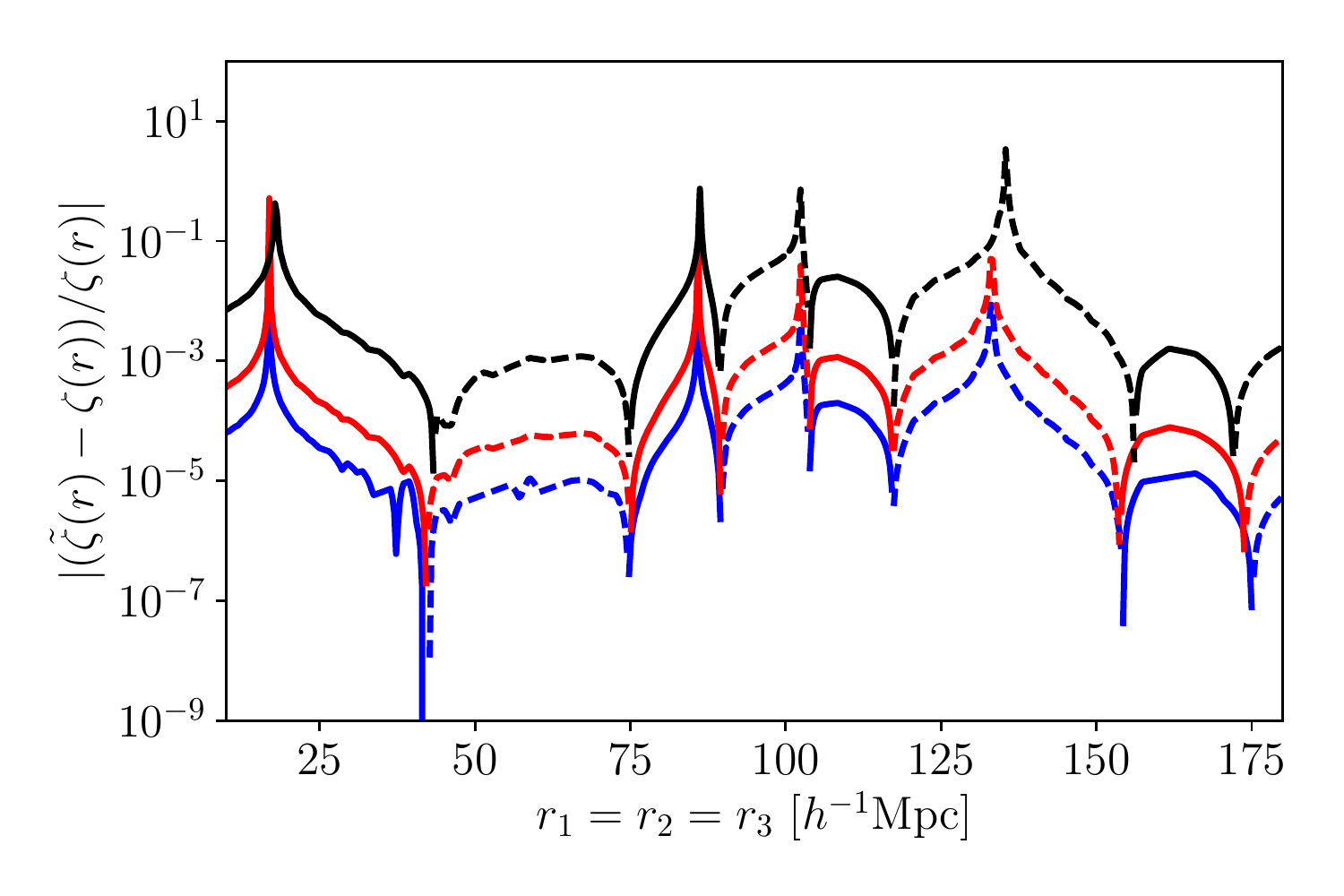}}
  \hfill
\subfloat{\includegraphics[width=0.48\textwidth]{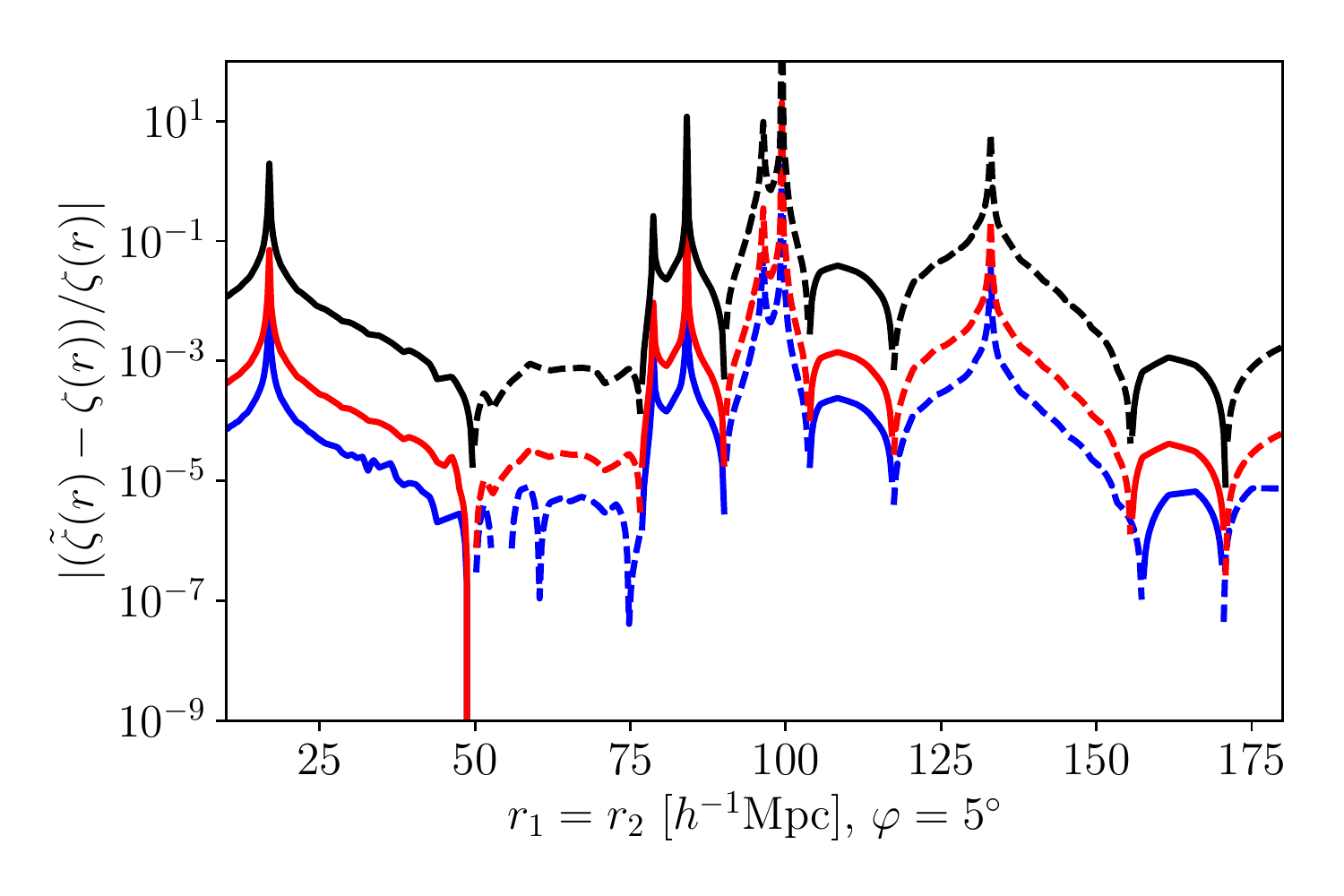}}
  \caption{Lensing deflection of 3PCF (blue: $z=0.5$, red: $z=1.0$, black: $z=10.0$).
\textit{Top}: Lensed - unlensed 3PCF. \textit{Bottom}: As proportion of unlensed 3PCF.  \textit{Left}:  Equilateral triangles. \textit{Right}: Squeezed triangles. Dashed lines indicate negative values.  } \label{fig:vz}
\end{figure}

\begin{figure}[t]
\centering
\subfloat{\includegraphics[width=0.5\textwidth]{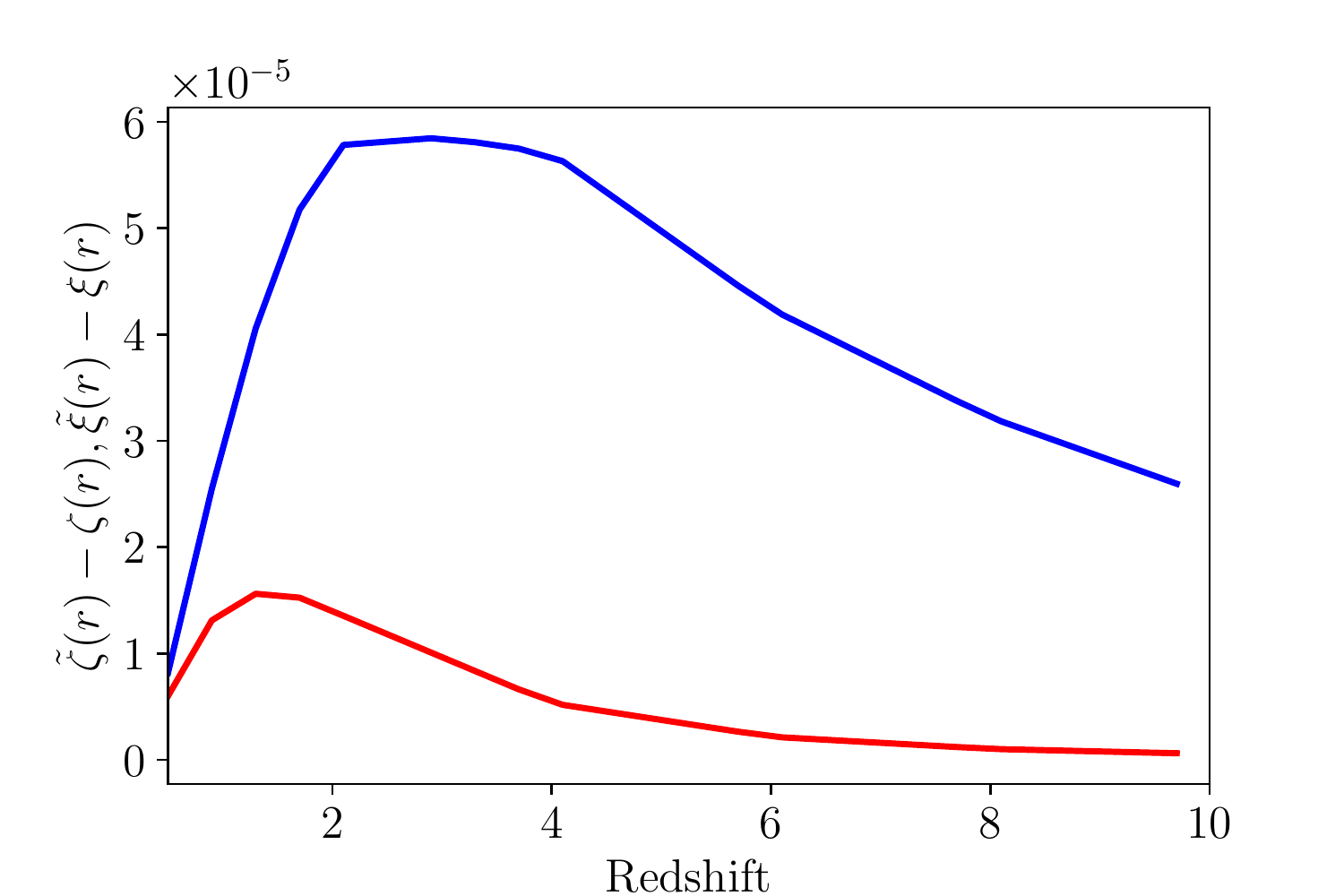}}
 \hfill
\subfloat{\includegraphics[width=0.5\textwidth]{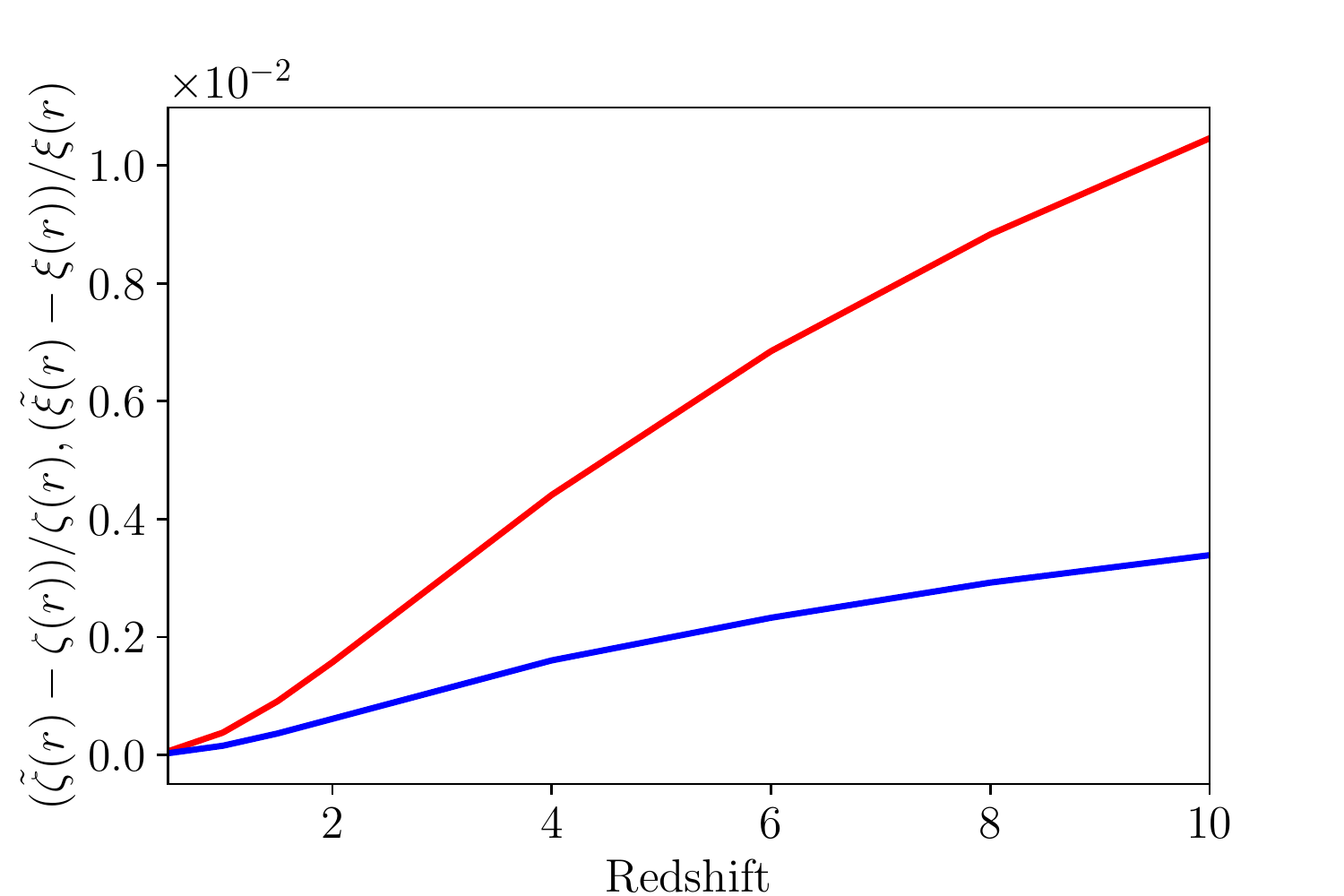}}
  \caption{Lensing deflection of the 2PCF (blue) and 3PCF (red) for equilateral triangles with sides \mbox{$r=10$ $h^{-1}$ Mpc}  as a function of redshift.
\textit{Left}: Absolute deflection. \textit{Right}: Relative deflection. } \label{fig:absvz}
\end{figure}

   Figure \ref{fig:vz} shows the lensing deflection and the relative deflection at different redshifts for equilateral and squeezed triangles. This demonstrates the significant effect of redshift on the lensing deflection, which has implications for the observability of the effect. The left panel of Figure \ref{fig:absvz} shows how the absolute lensing deflection varies with $z$ for equilateral triangles with sides of 10 $h^{-1}$ Mpc (chosen as illustrative of small scales, although in fact the relationship is similar at all scales). The absolute size of the deflection initially increases as $z$ increases up to about $z=2$, then decreases. This also occurs for the lensed 2PCF, also shown in Figure \ref{fig:absvz} for comparison, but in this case the highest deflection is at $z\sim3$. The shapes of these curves are due to the interplay between the lensing factors in Eqs. (\ref{eq:AB}) and (\ref{eq:ABC2*}) and the shapes of the correlation functions, through their derivatives.
 By contrast, as the right panel of Figure \ref{fig:absvz} shows, the lensing deflection as a proportion of the unlensed 3PCF and 2PCF increases monotonically with redshift.    Deflection decreases at higher redshifts but the correlation functions fall more rapidly. This can also be seen in Figure \ref{fig:vz}.   
  
\clearpage
\begin{figure}[t]
\centering
\text{             Equilateral}
\hspace{6cm}
\text{Squeezed}
\subfloat{\includegraphics[width=0.45\textwidth]{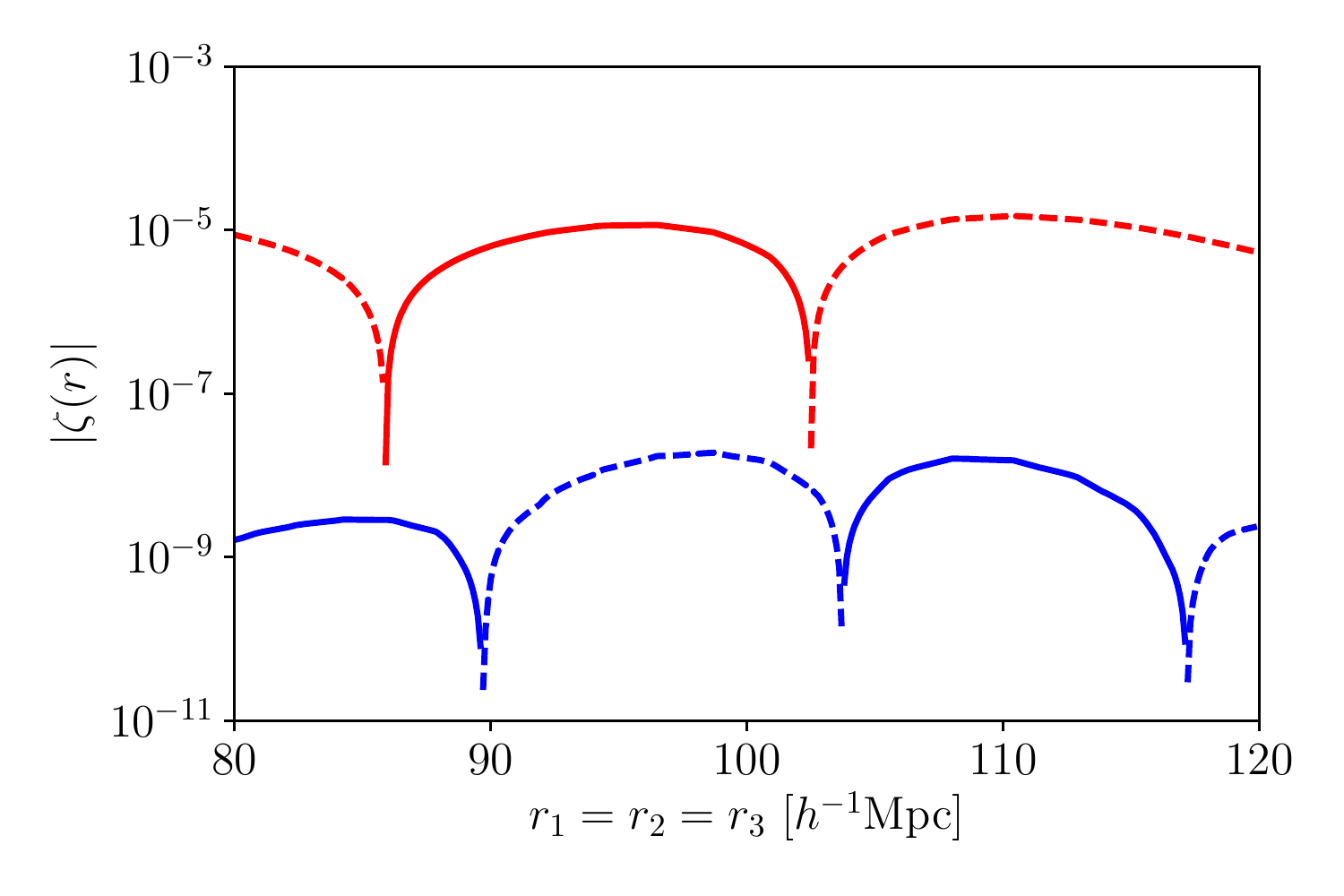}}
\hfill
\subfloat{\includegraphics[width=0.45\textwidth]{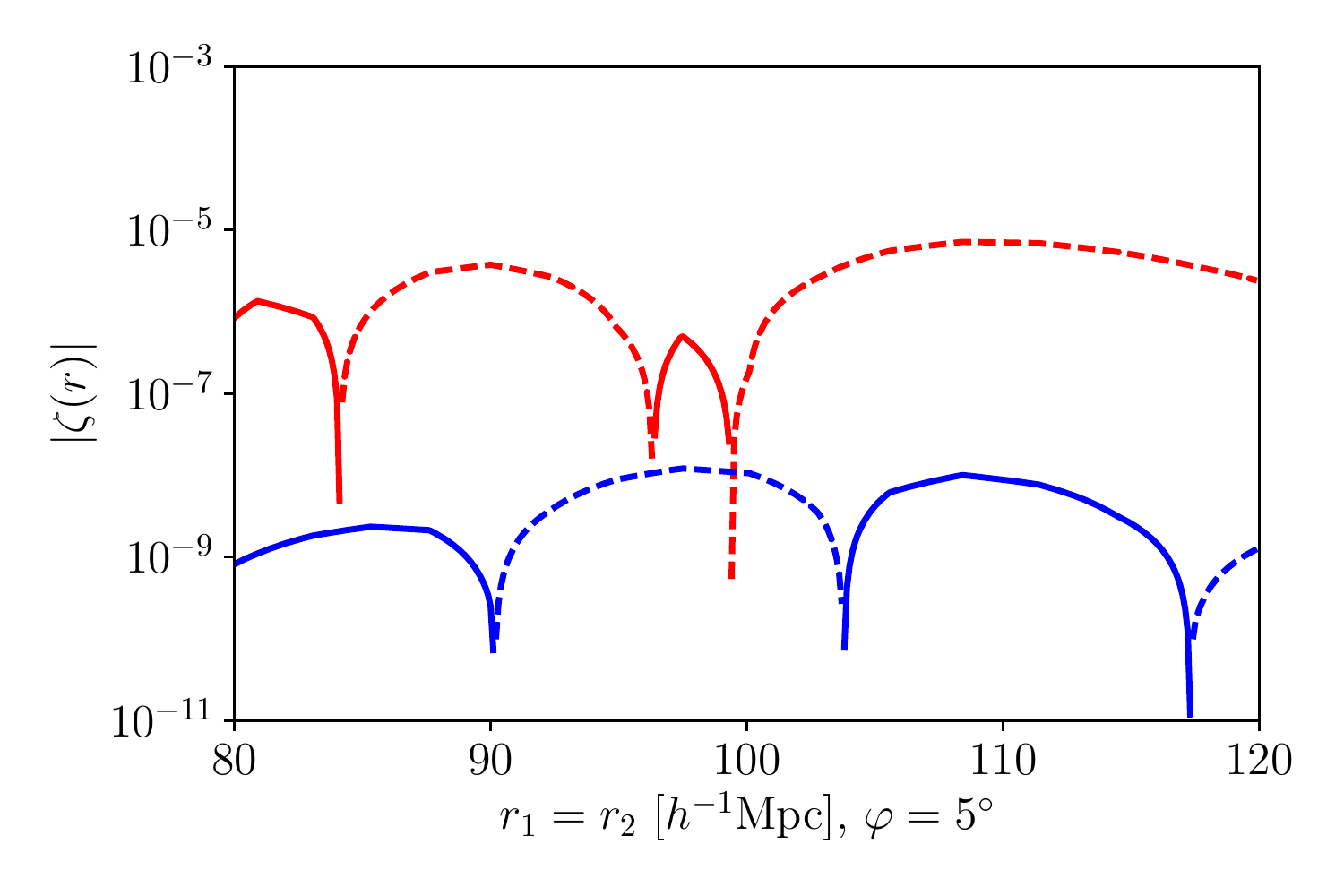}}   
\caption{Unlensed three-point correlation function (red) and lensing deflection (blue) near the BAO feature at $z=1$.
\textit{Left}: Equilateral triangles. \textit{Right}: Squeezed triangles. Dashed lines indicate negative values.} \label{fig:3PCF_BAO}
\end{figure}  
\begin{figure}[t]
\centering
\text{             Equilateral}
\hspace{6cm}
\text{Squeezed}
\subfloat{\includegraphics[width=0.48\textwidth]{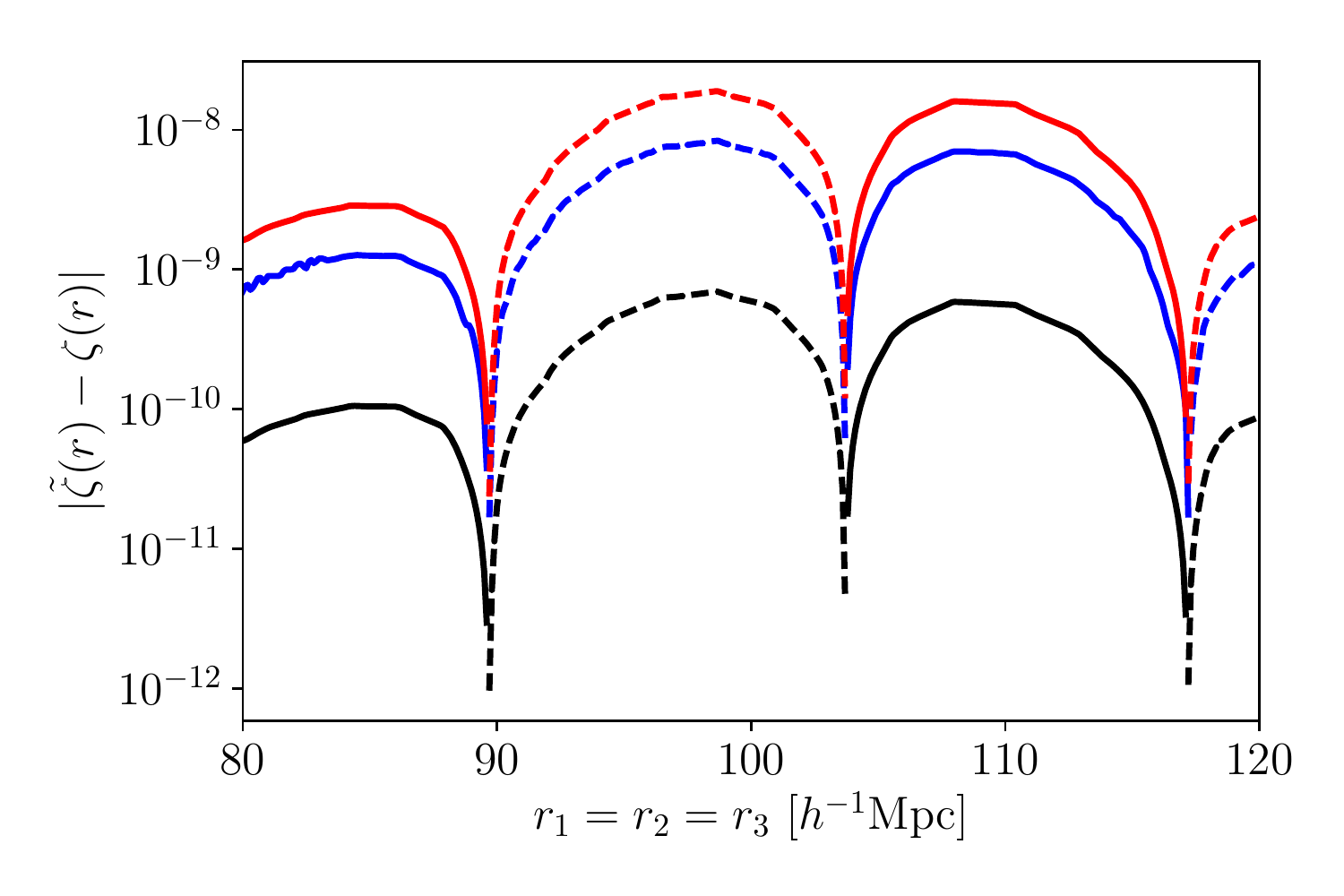}}
\hfill
\subfloat{\includegraphics[width=0.48\textwidth]{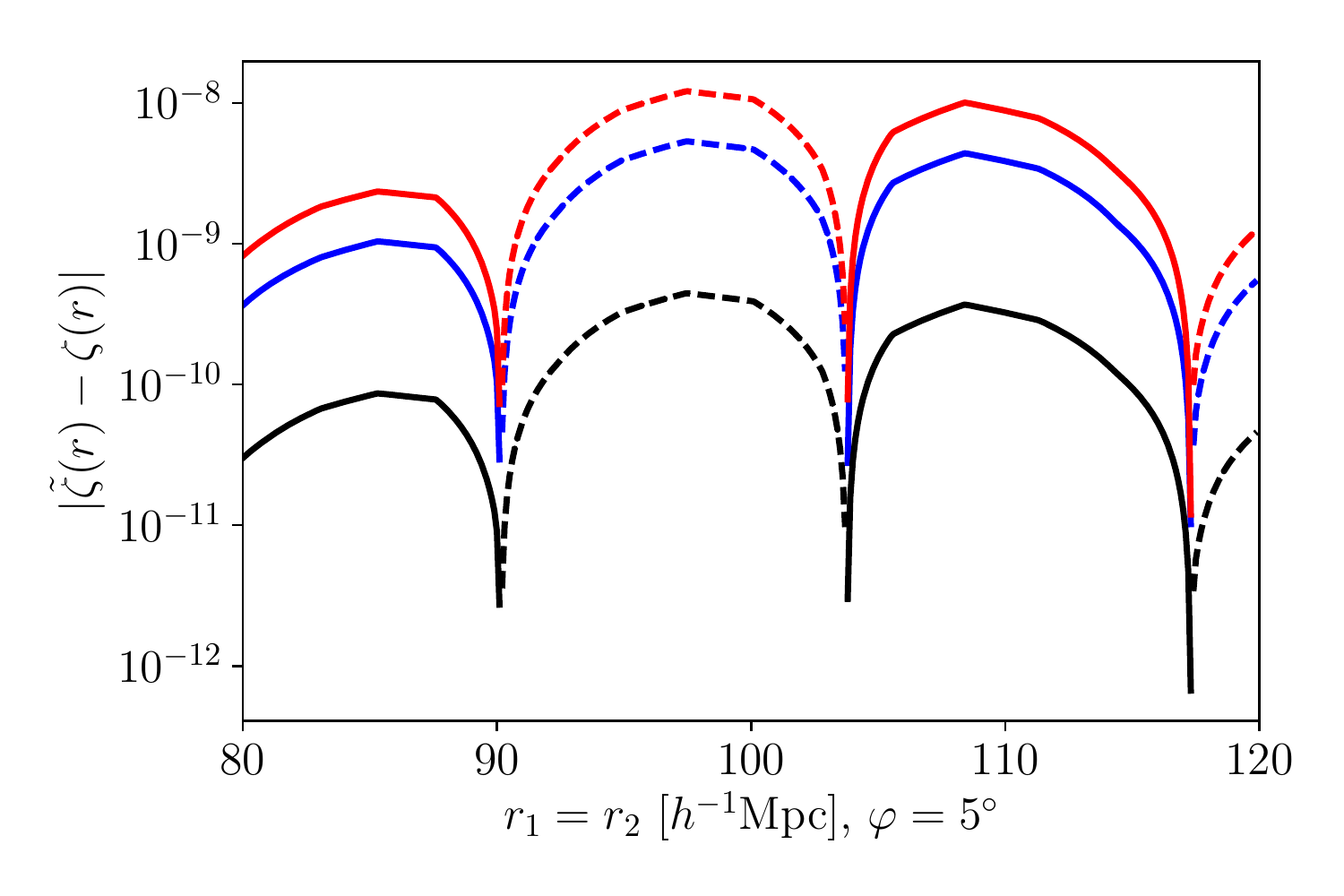}}
\vfill
\subfloat{\includegraphics[width=0.48\textwidth]{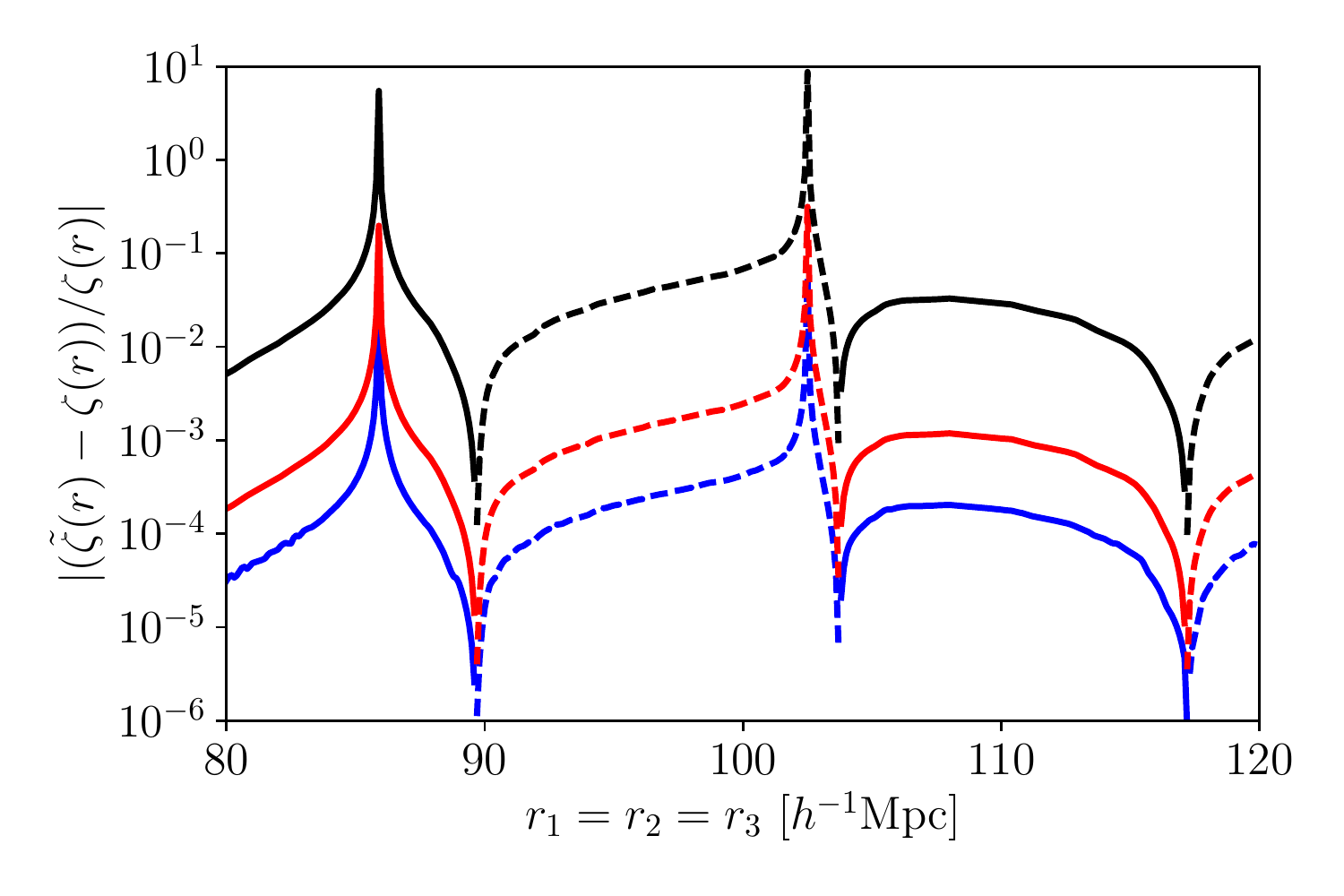}}
\hfill
\subfloat{\includegraphics[width=0.48\textwidth]{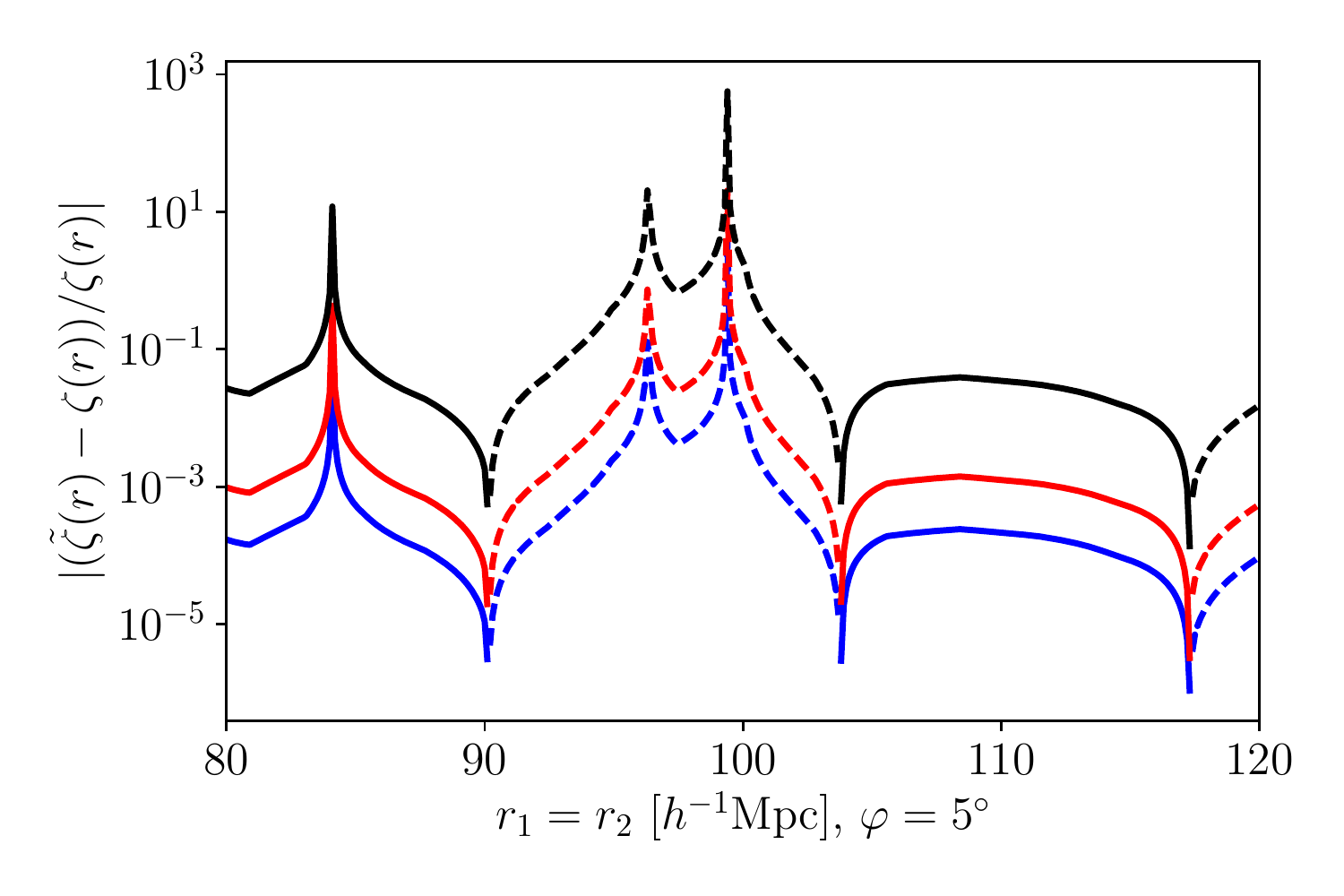}}
\caption{Lensing deflection of 3PCF near the BAO feature for different redshifts (blue: $z=0.5$, red: $z=1.0$, black: $z=10.0$).
\textit{Top}: Lensed - unlensed 3PCF. \textit{Bottom}: As proportion of unlensed 3PCF. \textit{Left}:  Equilateral triangles. \textit{Right}: Squeezed triangles. Dashed lines indicate negative values. } \label{fig:BAO}
\end{figure}
\clearpage
\subsection{BAO scale}\label{BAO}
Figure \ref{fig:3PCF_BAO} \lq{}zooms in\rq{} on the unlensed 3PCF at the BAO scale at $z=1$.  At this scale the 3PCF oscillates rapidly and vanishes at several points. This is particularly evident for squeezed triangles which display more structure.

Figure \ref{fig:BAO} shows the lensing deflection  and the relative deflection near the BAO feature at different redshifts.
Lensing deflection is more prominent at the BAO scale because the partial derivatives in Eq. (\ref{eq:ABC2*}) can be large.    Figures \ref{fig:3PCF_BAO} and \ref{fig:BAO} show that the lensing deflection smooths out  oscillations. At extrema of the 3PCF its first derivatives vanish and the lensing deflection depends on second derivatives. These are positive at  local minima, which means that the lensing deflection increases the 3PCF, and negative at  local maxima, decreasing the 3PCF.   At $z=1$ the peak-to-trough difference is smoothed by about 0.1 percent. This rises to around 2.3 percent at $z=10$.  

Since the unlensed 3PCF is zero at several values of $r$, the relative deflection becomes very large  in some regions.  However, we caution that the observability of the modification due to lensing depends on comparisons between the absolute (not relative) deflection and  statistical uncertainty on the 3PCF. We discuss this in the next section. 

\section{Observability of the lensing deflection }\label{obs}
The previous section shows that at the BAO scale the absolute value of the lensing deflection in the matter 3PCF is around $10^{-8}$  at $z=1$.  To assess whether a signal of this magnitude could be detected in current or future galaxy surveys, we assume that the uncertainty in survey measurements is entirely due to Poisson shot noise,  ignoring cosmic variance and other Gaussian and non-Gaussian errors which contribute to the full covariance.  Thus our error estimates are conservative and assume a minimum level of statistical error.    

To derive an estimate of the shot noise in the 3PCF we build on expressions for the shot noise in measurements of the bispectrum, for 3D fields \cite{RN73,RN79} and for projected fields \cite{RN72}. 

 Assuming the Gaussian limit,  the shot noise $\sigma^2(B)$ in  the Gaussian elements of the 3D bispectrum covariance can be estimated as \cite{RN73} 
\begin{align}
 	\sigma^2(B)&= \frac{ s_{123}}{V_sV_B\bar{n}^3}\ ,
\end{align}
where $V_s$ is the survey volume, $\bar{n}$ is the number density, $s_{123}=$ 1, 2 or 6 for general, isosceles or equilateral triangles respectively,  and $V_B$ quantifies the number of Fourier modes satisfying the triangle constraint
 $\mathbf{k}_1 + \mathbf{k}_2+ \mathbf{k}_3=0$.   It is  the integral over triangle side lengths $k_1, k_2,k_3$  of three spherical shells of width $\Delta k$. An analytical expression exists for $V_B$ \cite{RN79}:  
\begin{align}
V_B&=\int_{k_1}\mathrm{d}^3p \int_{k_2}\mathrm{d}^3q \int_{k_3}\mathrm{d}^3s\  \delta_\mathrm{D}(\mathbf{p}+\mathbf{q}+\mathbf{s})\\
&= 8\pi^2k_1k_2k_3(\Delta k)^3\ .
\end{align}

The quantity $V_B$ is a purely geometric measure, so similar reasoning applies to the 3PCF covariance. In this case we count modes satisfying the triangle constraint $\mathbf{r}_1 + \mathbf{r}_2+ \mathbf{r}_3=0$ within the bin width $\Delta r$ in real space to produce a quantity $V_Z$, analogous to $V_B$. 
Thus the  shot noise in the 3PCF covariance, $\sigma^2(\zeta)$, is given by
\begin{align}
\sigma^2(\zeta)&=\frac{(2\pi)^3s_{123}}{V_sV_Z\bar{n}^3}\\
&= \frac{6\pi}{V_sr^3(\Delta r)^3\bar{n}^3} \text{ for equilateral triangles.}\label{eq:error}
\end{align}

For the Euclid spectroscopic survey  $V_s\approx 100\  h^{-3}\text{Gpc}^3 $ and the expected number density of H$\alpha$ galaxies at $z=1$ is $\bar{n}\approx 1.7 \times 10^{-3}\ h^3\text{Mpc}^{-3}$  \cite{RN53}. We take $r=100\  h^{-1}\text{Mpc}$ (approximately the BAO scale, which is the scale of most interest) and ${\Delta r=10\  h^{-1}\text{Mpc}}$.  Inserting these values into Eq. (\ref{eq:error}) implies that  $\sigma^2(\zeta)$  is around $10^{-12}$ and $\sigma(\zeta) \sim 10^{-6}$. These order of magnitude estimates are consistent with estimates for the uncertainty in the  galaxy 3PCF in a survey similar to SDSS DR12  \cite{RN71}.  

 Thus, at the BAO scale and for equilateral triangles, the shot noise in a Euclid-like survey is greater than the deflection effect.      Since we have ignored several sources of error, in practice the errors will exceed the deflection effect by an even greater amount than calculated here.    
 
 The deflection is larger for squeezed triangles, but so too is the shot noise.  Deflection peaks at around $z=2$ (Figure \ref{fig:absvz}) but the number density of objects in Euclid-like surveys falls rapidly up to and beyond this redshift.  
 Other spectroscopic surveys, in particular DESI, will provide samples with similar number densities at $z<2$ and much sparser QSO samples beyond.  Photometric surveys like the LSST in principle observe deep tracer samples with high spatial densities, but the large line-of-sight uncertainties due to broadband photometric redshifts wash out small features like BAO signatures and the lensing deflection modification to the signal. 
Thus the deflection effect is too small to be detected by forthcoming galaxy and quasar surveys.  To put this into context, Ref. \cite{RN12} found the effect of  lensing magnification to be around $10^{-5}$ across a range of scales at $z=1$.  On the basis of fairly optimistic assumptions they considered this just detectable in planned surveys. 

\section{Conclusions}
We have derived an expression for the effect of lensing deflection on the three-point correlation function for three sources at the same comoving distance. The derivation is quite general: it could be applied to any physical observables and is based only on the assumptions that terms above second order in the lensing deflection can be neglected and that the observables are not correlated with the lensing deflection field. We do not assume that the lensing deflection field is Gaussian. 

 The resulting expression, given by Eq. (\ref{eq:ABC2*}), shows that the lensing deflection depends on partial derivatives of the unlensed 3PCF. This causes the 3PCF to be smoothed by lensing, but also makes the effect large when the 3PCF is rapidly varying.  If the 3PCF is approximately a power law (which will often be the case), each term of the lensing deflection  in Eq. (\ref{eq:ABC2*}) behaves like $\langle ABC \rangle/r^2$. Thus as a proportion of the unlensed 3PCF, the deflection effect is highest for sources which are close to each other and decreases as the source separation increases. At all scales the relative effect increases monotonically with redshift. 

We have calculated the size of this effect for the matter density contrast and show that lensing deflection is around $10^{-8}$  at $z=1$. We have confirmed that the effect is highest at small scales ($r<20 \ h^{-1}$ Mpc), and  is also especially noticeable around the BAO feature where it smooths out the peaks and troughs, reducing the amplitude of the oscillations. This could potentially affect the use of the 3PCF for cosmological parameter estimation.   The deflection is greatest at around $z=2$ but as a proportion of the unlensed 3PCF the deflection increases with redshift, reaching $10^{-2}$ at $z=10$.  A similar results holds for the 2PCF. Here the peak deflection is greater but occurs at higher redshift. 

The effect on the 3PCF is too small to be detected in forthcoming surveys such as Euclid or DESI. Detections would require  much higher number densities of galaxies or quasars at redshifts $z\sim2$ where the deflection is greatest.   While we cannot directly compare our findings with results obtained in Fourier space which were also considered undetectable \cite{RN29}, we expect them to be of the same order of magnitude. 
\acknowledgments {BJ acknowledges support by an STFC Ernest Rutherford Fellowship, grant reference ST/J004421/1.
HVP was partially supported by the European Research Council (ERC) under the European Community's Seventh Framework Programme (FP7/2007-2013)/ERC grant agreement number 306478- CosmicDawn.} This work was partially enabled by funding from the UCL Cosmoparticle Initiative.
\renewcommand*{\refname}{} 
\section
*{References}   
\bibliography{3PCF_refs}
\vspace{1cm}
\appendix
\setlength{\parindent}{0pt}

\Large\textbf{Appendices}
\normalsize

\renewcommand{\theequation}{\Alph{section}.\arabic{equation}}

\section {Effect of lensing on the three-point correlation function}\label{L3PCF}

Consider three physical observables $A(\mathbf{x}_a)$, $ B(\mathbf{x}_b)$  and $C(\mathbf{x}_c)$ observed at points $\mathbf{x}_a$, $\mathbf{x}_b$  and $\mathbf{x}_c$. The corresponding observed (lensed) values are $ \tilde{A}(\mathbf{x}_a)$, $\tilde{B}(\mathbf{x}_b)$  and $\tilde{C}(\mathbf{x}_c)$.
Then if $\boldsymbol{\lambda}_a$ is the lensing deflection vector, $\tilde{A}(\mathbf{x}_a)$ and $A(\mathbf{x}_a+\boldsymbol{\lambda}_a)$ are related by Eq.(\ref{eq:A}), and similarly for $B$ and $C$.

The unlensed 3PCF is
\begin{align}
\langle A(\mathbf{x}_a)B(\mathbf{x}_b) C(\mathbf{x}_c)\rangle&=\int\frac{\mathrm{d}^3k_1}{(2\pi)^3}\mathrm{e}^{\mathrm{i}\mathbf{k}_1\cdot\mathbf{x}_a}\int\frac{\mathrm{d}^3k_2}{(2\pi)^3}\mathrm{e}^{\mathrm{i}\mathbf{k}_2\cdot\mathbf{x}_b}
	\int\frac{\mathrm{d}^3k_3}{(2\pi)^3}\mathrm{e}^{\mathrm{i}\mathbf{k}_3\cdot\mathbf{x}_c}\notag\\
&\quad\quad\quad\quad\times (2\pi)^3  B_{ABC}(\mathbf{k}_1,\mathbf{k}_2,\mathbf{k}_3)\delta_\text{D}(\mathbf{k}_1+\mathbf{k}_2+\mathbf{k}_3)\ ,\label{eq:u3PCF}
\end{align}
 and the lensed 3PCF is
\begin{align}
	\langle\tilde{A}(\mathbf{x}_a)\tilde{B}(\mathbf{x}_b)\tilde{C}(\mathbf{x}_c)\rangle
	& =\langle A(\mathbf{x}_a+\boldsymbol{\lambda_a})B(\mathbf{x}_b+\boldsymbol{\lambda_b}) C(\mathbf{x}_c+\boldsymbol{\lambda_c})\rangle\\
	&=\int\frac{\mathrm{d}^3k_1}{(2\pi)^3}\mathrm{e}^{\mathrm{i}\mathbf{k}_1\cdot(\mathbf{x}_a+\boldsymbol{\lambda_a})}\int\frac{\mathrm{d}^3k_2}{(2\pi)^3}\mathrm{e}^{\mathrm{i}\mathbf{k}_2\cdot(\mathbf{x}_b+\boldsymbol{\lambda_b})}
	\int\frac{\mathrm{d}^3k_3}{(2\pi)^3}\mathrm{e}^{\mathrm{i}\mathbf{k}_3\cdot(\mathbf{x}_c+\boldsymbol{\lambda_c})}\notag\\
	&\quad\quad\quad\quad\times (2\pi)^3 B_{ABC}(\mathbf{k}_1,\mathbf{k}_2,\mathbf{k}_3)\delta_\text{D}(\mathbf{k}_1+\mathbf{k}_2+\mathbf{k}_3)\ ,\label{eq:C4}
\end{align}
 where $\delta_\text{D}$ is the Dirac delta function and  $B_{ABC}(\mathbf{k}_1,\mathbf{k}_2,\mathbf{k}_3)$ is the bispectrum of the three observables, defined as:
\begin{align}
\langle A(\mathbf{x}_a)B(\mathbf{x}_b) C(\mathbf{x}_c)\rangle &= (2 \pi)^3 \delta_\mathrm{D}(\mathbf{k}_1 +\mathbf{k}_2+\mathbf{k}_3)B_{ABC}(\mathbf{k}_1,\mathbf{k}_2,\mathbf{k}_3)\ .
\end{align} 
We now derive an expression for the lensed 3PCF in terms of the unlensed 3PCF and a lensing deflection term which we denote $\langle ABC \rangle_2$.

 Dropping the arguments of $ \tilde{A}(\mathbf{x}_a)$ etc for simplicity and making use of the delta function in Eq. (\ref{eq:C4})  leads to
\begin{align}
	\langle\tilde{A}\tilde{B}\tilde{C}\rangle&=\int\frac{\mathrm{d}^3k_1}{(2\pi)^3}\int\frac{\mathrm{d}^3k_2}{(2\pi)^3}\langle \mathrm{e}^{\mathrm{i}\mathbf{k}_1\cdot(\mathbf{x}_a+\boldsymbol{\lambda_a}-\mathbf{x}_c-\boldsymbol{\lambda_c})} 
	\mathrm{e}^{\mathrm{i}\mathbf{k}_2\cdot(\mathbf{x}_b+\boldsymbol{\lambda_b}-\mathbf{x}_c-\boldsymbol{\lambda_c})}\rangle B_{ABC}(\mathbf{k}_1,\mathbf{k}_2, -\mathbf{k}_1-\mathbf{k}_2)\notag\\
	&=\int\frac{\mathrm{d}^3k_1}{(2\pi)^3}\int\frac{\mathrm{d}^3k_2}{(2\pi)^3}B_{ABC}(\mathbf{k}_1,\mathbf{k}_2, -\mathbf{k}_1-\mathbf{k}_2)\mathrm{e}^{\mathrm{i}\mathbf{k}_1\cdot(\mathbf{x}_a-\mathbf{x}_c)}\mathrm{e}^{\mathrm{i}\mathbf{k}_2\cdot(\mathbf{x}_b-\mathbf{x}_c)}\langle\mathrm{e}^{\mathrm{i}\mathbf{k}_1\cdot(\boldsymbol{\lambda}_a
-\boldsymbol{\lambda}_c)}\mathrm{e}^{\mathrm{i}\mathbf{k}_2\cdot(\boldsymbol{\lambda}_b-\boldsymbol{\lambda}_c)}\rangle\  ,
\end{align}
assuming the observables are not correlated with the lensing deflection field.

We expand the exponential factors in the expectation value  up to second order in ${k}$ to get
\begin{align}
\langle\tilde{A}\tilde{B}\tilde{C}\rangle&\approx\int\frac{\mathrm{d}^3k_1}{(2\pi)^3}\int\frac{\mathrm{d}^3k_2}{(2\pi)^3}B_{ABC}(\mathbf{k}_1,\mathbf{k}_2, -\mathbf{k}_1-\mathbf{k}_2)\mathrm{e}^{\mathrm{i}\mathbf{k}_1\cdot(\mathbf{x}_a-\mathbf{x}_c)}\mathrm{e}^{\mathrm{i}\mathbf{k}_2\cdot(\mathbf{x}_b-\mathbf{x}_c)}\notag\\
	& \hspace{3cm} \times  \langle [1+\mathrm{i}\mathbf{k}_1\cdot(\boldsymbol{\lambda}_a-\boldsymbol{\lambda}_c) -\frac{1}{2}\mathbf{k}_1\cdot(\boldsymbol{\lambda}_a-\boldsymbol{\lambda}_c)\mathbf{k}_1\cdot(\boldsymbol{\lambda}_a-\boldsymbol{\lambda}_c)] \notag\\
	&\hspace{3cm}\times [1+\mathrm{i}\mathbf{k}_2\cdot(\boldsymbol{\lambda}_b-\boldsymbol{\lambda}_c) -\frac{1}{2}\mathbf{k}_2\cdot(\boldsymbol{\lambda}_b-\boldsymbol{\lambda}_c)\mathbf{k}_2\cdot(\boldsymbol{\lambda}_b-\boldsymbol{\lambda}_c)] \rangle\ .
\end{align}
The zeroth order term is the unlensed 3PCF, $\langle ABC \rangle$. The terms like $\mathrm{i}\mathbf{k}_2\cdot(\boldsymbol{\lambda}_b-\boldsymbol{\lambda}_c)$ are zero because the expectation value of the deflection field is zero.  So we have
\begin{align}
\langle\tilde{A}\tilde{B}\tilde{C}\rangle&\approx\langle ABC \rangle +\int\frac{\mathrm{d}^3k_1}{(2\pi)^3}\int\frac{\mathrm{d}^3k_2}{(2\pi)^3}B_{ABC}(\mathbf{k}_1,\mathbf{k}_2, -\mathbf{k}_1-\mathbf{k}_2)\mathrm{e}^{\mathrm{i}\mathbf{k}_1\cdot(\mathbf{x}_a-\mathbf{x}_c)}\mathrm{e}^{\mathrm{i}\mathbf{k}_2\cdot(\mathbf{x}_b-\mathbf{x}_c)}\notag\\
&\quad\quad\times\langle-\frac{1}{2}\mathbf{k}_1\cdot(\boldsymbol{\lambda}_a-\boldsymbol{\lambda}_c)\mathbf{k}_1\cdot(\boldsymbol{\lambda}_a-\boldsymbol{\lambda}_c) -\frac{1}{2}\mathbf{k}_2\cdot(\boldsymbol{\lambda}_b-\boldsymbol{\lambda}_c)\mathbf{k}_2\cdot(\boldsymbol{\lambda}_b-\boldsymbol{\lambda}_c)\notag\\
&\hspace{4cm}-\mathbf{k}_1\cdot(\boldsymbol{\lambda}_a-\boldsymbol{\lambda}_c)\mathbf{k}_2\cdot(\boldsymbol{\lambda}_b-\boldsymbol{\lambda}_c)\rangle\label{eq:ABCcalc}\\
&\equiv \langle ABC\rangle + \langle ABC \rangle_2\ .
\end{align}
The expectation value in Eq. (\ref{eq:ABCcalc})  can be written 
\begin{align}
&\langle-\frac{1}{2}\mathbf{k}_1\cdot(\boldsymbol{\lambda}_a-\boldsymbol{\lambda}_c)\mathbf{k}_1\cdot(\boldsymbol{\lambda}_a-\boldsymbol{\lambda}_c) -\frac{1}{2}\mathbf{k}_2\cdot(\boldsymbol{\lambda}_b-\boldsymbol{\lambda}_c)\mathbf{k}_2\cdot(\boldsymbol{\lambda}_b-\boldsymbol{\lambda}_c)
-\mathbf{k}_1\cdot(\boldsymbol{\lambda}_a-\boldsymbol{\lambda}_c)\mathbf{k}_2\cdot(\boldsymbol{\lambda}_b-\boldsymbol{\lambda}_c)\rangle\notag\\
&=-\frac{1}{2}\big[k_{1i}k_{1j}[ \langle \lambda_a^i\lambda_a^j\rangle- 2\langle\lambda_a^i\lambda_c^j \rangle+\langle \lambda_c^i\lambda_c^j  \rangle]
 +k_{2i}k_{2j}[ \langle \lambda_b^i\lambda_b^j\rangle- 2\langle\lambda_b^i\lambda_c^j \rangle+\langle \lambda_c^i\lambda_c^j  \rangle]\big]\notag\\
&\hspace{5cm}-k_{1i}k_{2j} [\langle \lambda_a^i\lambda_b^j\rangle -\langle \lambda_a^i\lambda_c^j\rangle - \langle \lambda_b^i\lambda_c^j\rangle+\langle\lambda_c^i\lambda_c^j  \rangle]\ , \label{eq:corr}
\end{align}
where summation over $i$ and $j$ is implied.
So the lensing deflection term in Eq. (\ref{eq:ABCcalc}) becomes
\begin{align}
\langle ABC \rangle_2&\approx\int\frac{\mathrm{d}^3k_1}{(2\pi)^3}\int\frac{\mathrm{d}^3k_2}{(2\pi)^3}B_{ABC}(\mathbf{k}_1,\mathbf{k}_2, -\mathbf{k}_1-\mathbf{k}_2)\mathrm{e}^{\mathrm{i}\mathbf{k}_1\cdot(\mathbf{x}_a-\mathbf{x}_c)}\mathrm{e}^{\mathrm{i}\mathbf{k}_2\cdot(\mathbf{x}_b-\mathbf{x}_c)}\notag\\
&\quad\quad\times\bigg[-\frac{1}{2}\big[k_{1i}k_{1j}[ \langle \lambda_a^i\lambda_a^j\rangle- 2\langle\lambda_a^i\lambda_c^j \rangle+\langle \lambda_c^i\lambda_c^j  \rangle]
 +k_{2i}k_{2j}[ \langle \lambda_b^i\lambda_b^j\rangle- 2\langle\lambda_b^i\lambda_c^j \rangle+\langle \lambda_c^i\lambda_c^j  \rangle]\big]\notag\\
&\quad\quad\quad\quad-k_{1i}k_{2j} [\langle \lambda_a^i\lambda_b^j\rangle -\langle \lambda_a^i\lambda_c^j\rangle - \langle \lambda_b^i\lambda_c^j\rangle+\langle\lambda_c^i\lambda_c^j  \rangle]\bigg]\ .
\end{align}
The deflection vectors do not depend on the integration variables and can be taken out of the integrals to give
\begin{align}
\langle ABC \rangle_2&\approx-\frac{1}{2}[ \langle \lambda_a^i\lambda_a^j\rangle- 2\langle\lambda_a^i\lambda_c^j \rangle+\langle \lambda_c^i\lambda_c^j  \rangle]
\int\frac{\mathrm{d}^3k_1}{(2\pi)^3}\int\frac{\mathrm{d}^3k_2}{(2\pi)^3}k_{1i}k_{1j}B_{ABC}(\mathbf{k}_1,\mathbf{k}_2, -\mathbf{k}_1-\mathbf{k}_2)\notag\\
&\hspace{8cm}\times  \mathrm{e}^{\mathrm{i}\mathbf{k}_1\cdot(\mathbf{x}_a-\mathbf{x}_c)}\mathrm{e}^{\mathrm{i}\mathbf{k}_2\cdot(\mathbf{x}_b-\mathbf{x}_c)}\notag\\
&\quad                                                        -\frac{1}{2}[ \langle \lambda_b^i\lambda_b^j\rangle- 2\langle\lambda_b^i\lambda_c^j \rangle+\langle \lambda_c^i\lambda_c^j  \rangle] 
\int\frac{\mathrm{d}^3k_1}{(2\pi)^3}\int\frac{\mathrm{d}^3k_2}{(2\pi)^3}k_{2i}k_{2j}B_{ABC}(\mathbf{k}_1,\mathbf{k}_2, -\mathbf{k}_1-\mathbf{k}_2)\notag\\
&\hspace{8cm}\times  \mathrm{e}^{\mathrm{i}\mathbf{k}_1\cdot(\mathbf{x}_a-\mathbf{x}_c)}\mathrm{e}^{\mathrm{i}\mathbf{k}_2\cdot(\mathbf{x}_b-\mathbf{x}_c)}\notag\\
&\quad-[\langle \lambda_a^i\lambda_b^j\rangle -\langle \lambda_a^i\lambda_c^j\rangle - \langle \lambda_b^i\lambda_c^j\rangle+\langle\lambda_c^i\lambda_c^j  \rangle]
\int\frac{\mathrm{d}^3k_1}{(2\pi)^3}\int\frac{\mathrm{d}^3k_2}{(2\pi)^3}k_{1i}k_{2j}B_{ABC}(\mathbf{k}_1,\mathbf{k}_2, -\mathbf{k}_1-\mathbf{k}_2)\notag\\
&\hspace{8cm}\times \mathrm{e}^{\mathrm{i}\mathbf{k}_1\cdot(\mathbf{x}_a-\mathbf{x}_c)}\mathrm{e}^{\mathrm{i}\mathbf{k}_2\cdot(\mathbf{x}_b-\mathbf{x}_c)}\label{eq:ABClens}\ .
\end{align}
 Now define the correlators as in Eq. (\ref{eq:Z}). Then the final correlator term in Eq. (\ref{eq:corr}) can be written as 
\begin {align}
	 \langle\lambda_a^i\lambda_b^j\rangle-\langle\lambda_a^i\lambda_c^j\rangle-\langle\lambda_b^i\lambda_c^j\rangle+\langle\lambda_c^i\lambda_c^j\rangle
	&= -\left(\frac{\langle\lambda_a^i\lambda_a^j\rangle+\langle\lambda_b^i\lambda_b^j\rangle}{2}-\langle\lambda_a^i\lambda_b^j\rangle\right)+ \left(\frac{\langle\lambda_a^i\lambda_a^j\rangle+\langle\lambda_b^i\lambda_b^j\rangle}{2}\right)\notag\\
	&\quad+\left(\frac{\langle\lambda_a^i\lambda_a^j\rangle+\langle\lambda_c^i\lambda_c^j\rangle}{2}-\langle\lambda_a^i\lambda_c^j\rangle\right)-\left(\frac{\langle\lambda_a^i\lambda_a^j\rangle+\langle\lambda_c^i\lambda_c^j\rangle}{2}\right)\notag\\
	&\quad+\left(\frac{\langle\lambda_b^i\lambda_b^j\rangle+\langle\lambda_c^i\lambda_c^j\rangle}{2}-\langle\lambda_b^i\lambda_c^j\rangle\right)- \left(\frac{\langle\lambda_b^i\lambda_b^j\rangle+\langle\lambda_c^i\lambda_c^j\rangle}{2}\right)\notag\\
	&\quad +\langle\lambda_c^i\lambda_c^j\rangle\notag\\
	&= Z_{ac}^{ij}+Z_{bc}^{ij}-Z_{ab}^{ij}\ .\\
\notag
\end{align}
So the lensing deflection is given by
\begin{align}
\langle ABC \rangle_2&\approx-Z_{ac}^{ij}
\int\frac{\mathrm{d}^3k_1}{(2\pi)^3}\int\frac{\mathrm{d}^3k_2}{(2\pi)^3}k_{1i}k_{1j}B_{ABC}(\mathbf{k}_1,\mathbf{k}_2, -\mathbf{k}_1-\mathbf{k}_2)\notag\\
&\hspace{8cm}\times  \mathrm{e}^{\mathrm{i}\mathbf{k}_1\cdot(\mathbf{x}_a-\mathbf{x}_c)} \mathrm{e}^{\mathrm{i}\mathbf{k}_2\cdot(\mathbf{x}_b-\mathbf{x}_c)}\notag\\
&\quad                                                     -Z_{bc}^{ij}
\int\frac{\mathrm{d}^3k_1}{(2\pi)^3}\int\frac{\mathrm{d}^3k_2}{(2\pi)^3}k_{2i}k_{2j}B_{ABC}(\mathbf{k}_1,\mathbf{k}_2, -\mathbf{k}_1-\mathbf{k}_2)\notag\\
&\hspace{8cm}\times   \mathrm{e}^{\mathrm{i}\mathbf{k}_1\cdot(\mathbf{x}_a-\mathbf{x}_c)} \mathrm{e}^{\mathrm{i}\mathbf{k}_2\cdot(\mathbf{x}_b-\mathbf{x}_c)}\notag\\
&\quad-[Z_{ac}^{ij}+Z_{bc}^{ij}-Z_{ab}^{ij}]
\int\frac{\mathrm{d}^3k_1}{(2\pi)^3}\int\frac{\mathrm{d}^3k_2}{(2\pi)^3}k_{1i}k_{2j}B_{ABC}(\mathbf{k}_1,\mathbf{k}_2, -\mathbf{k}_1-\mathbf{k}_2)\notag\\
&\hspace{8cm}\times  \mathrm{e}^{\mathrm{i}\mathbf{k}_1\cdot(\mathbf{x}_a-\mathbf{x}_c)} \mathrm{e}^{\mathrm{i}\mathbf{k}_2\cdot(\mathbf{x}_b-\mathbf{x}_c)}\label{eq:ABClens}\ .
\end{align}
We now define $\mathbf{r}_b$ as the observed distance between $\mathbf{x}_a$ and $\mathbf{x}_c$, and $\mathbf{r}_a$ as the observed distance between $\mathbf{x}_b$ and $\mathbf{x}_c$, so that $\mathbf{r}_b\equiv(\mathbf{x}_a-\mathbf{x}_c)$  and  $\mathbf{r}_a\equiv(\mathbf{x}_b-\mathbf{x}_c)$ (see Figure \ref{fig:coords3}).  Then from Eq. (\ref{eq:u3PCF}) 
\begin{align}
\langle A(\mathbf{x}_a)B(\mathbf{x}_b) C(\mathbf{x}_c)\rangle&=\int\frac{\mathrm{d}^3k_1}{(2\pi)^3}\mathrm{e}^{\mathrm{i}\mathbf{k}_1\cdot(\mathbf{r}_b+\mathbf{x}_c)}\int\frac{\mathrm{d}^3k_2}{(2\pi)^3}\mathrm{e}^{\mathrm{i}\mathbf{k}_2\cdot(\mathbf{r}_a+\mathbf{x}_c)}
	\int\frac{\mathrm{d}^3k_3}{(2\pi)^3}\mathrm{e}^{\mathrm{i}\mathbf{k}_3\cdot\mathbf{x}_c}\notag\\
&\quad\quad\quad\quad\times (2\pi)^3  B_{ABC}(\mathbf{k}_1,\mathbf{k}_2,\mathbf{k}_3)\delta_\text{D}(\mathbf{k}_1+\mathbf{k}_2+\mathbf{k}_3)\ ,
\end{align}
 and so partial derivatives with respect to components of the triangle sides can be written as, for example,
\begin{align}
 \psderiv{\langle ABC\rangle}{r_{bi}}{r_{bj}}&=  -k_{1i}k_{1j}\langle ABC\rangle\ .
\end{align} 
\\
It follows that the lensing deflection, $\langle ABC\rangle_2$, can be expressed in terms of partial derivatives of the unlensed 3PCF, $\langle ABC \rangle$:
\begin{align}
	\langle ABC\rangle_2&= \psderiv{\langle ABC\rangle}{r_{bi}}{r_{bj}}Z_{ac}^{ij}\notag\\
	&\quad+ \psderiv{\langle ABC\rangle}{r_{ai}}{r_{aj}}Z_{bc}^{ij}\notag\\
	&\quad+\psderiv{\langle ABC\rangle}{r_{ai}}{r_{bj}}[Z_{ac}^{ij}+Z_{bc}^{ij}-Z_{ab}^{ij}]\label{eq:ABC2pd1}\ .
\end{align}
\\
Following ref. \cite{RN1} we now make the simplifying assumption that all three sources are at the same comoving distance.  Without loss of generality coordinates can be chosen as in Figure \ref{fig:coords3}. This means that the only elements of the distortion correlators $Z^{ij}$ which we need to consider are those with $i=j=1$ and $i=j=2$. These represent deflections in two orthogonal directions in the plane of the sky. With these coordinates  Eq. (\ref{eq:ABC2pd1}) becomes
\begin{align}
	\langle ABC\rangle_2&= \psderiva{\langle ABC\rangle}{r_{b1}}Z_{ac}^{11}+ \psderiva{\langle ABC\rangle}{r_{b2}}Z_{ac}^{22}\notag\\
	&\quad+ \psderiva{\langle ABC\rangle}{r_{a1}}Z_{bc}^{11}+\psderiva{\langle ABC\rangle}{r_{a2}}Z_{bc}^{22}\notag\\
	&\quad+\psderiv{\langle ABC\rangle}{r_{a1}}{r_{b1}}[Z_{ac}^{11}+Z_{bc}^{11}-Z_{ab}^{11}]\notag\\
	&\quad+\psderiv{\langle ABC\rangle}{r_{a2}}{r_{b2}}[Z_{ac}^{22}+Z_{bc}^{22}-Z_{ab}^{22}]\\
	&=\psderiva{\langle ABC\rangle}{r_{b1}}Z_{ac}^{11}+\psderiva{\langle ABC\rangle}{r_{a1}}Z_{bc}^{11}+\psderiva{\langle ABC\rangle}{r_{a2}}Z_{bc}^{22}\notag\\
           &\quad+\psderiv{\langle ABC\rangle}{r_{a1}}{r_{b1}}[Z_{ac}^{11}+Z_{bc}^{11}-Z_{ab}^{11}]\label{eq:ABC2pd2}
\end{align}
because all derivatives with respect to $r_{b2}$ are zero through the choice of coordinates.

We next express these derivatives in terms of the distances ${r_a}$ and ${r_b}$ and the angle $\varphi$ between $\mathbf{r}_a$ and $\mathbf{r}_b$.  We have $r_{b1}\equiv r_b$ so partial derivatives with respect to $r_{b1}$ are straightforward.  We now derive the other partial derivatives which appear in Eq. (\ref{eq:ABC2pd2}). For brevity we write $\langle ABC\rangle$ as $f\equiv f(r_a,r_b,\varphi)$.
\\\\
\textbf{1. Partial derivative with respect to $r_{a1}$.}
\begin{align}
	\pderiv{f(r_a,r_b,\varphi)}{r_{a_{1}}}
        &= \pderiv{f}{r_a}\pderiv{r_a}{r_{a_{1}}}+ \pderiv{f}{r_b}\pderiv{r_b}{r_{a_{1}}}+ \pderiv{f}{\varphi}\pderiv{\varphi}{r_{a_{1}}}\notag\\
	 &=\cos\varphi\pderiv{f}{r_a} -\frac{\sin \varphi}{r_a}\pderiv{f}{\varphi}\ .    
 \end{align}

This uses  $r_a = (r_{a1}^2+ r_{a2}^2)^{\nicefrac{1}{2}}$ which means $\pderiv{r_a}{r_{a1}} = \frac{r_{a1}}{(r_{a1}^2+ r_{a2}^2)^{\nicefrac{1}{2}}}$ \mbox{$= \nicefrac{r_{a1}}{r_a} = \cos \varphi$} ,    \\
and $\varphi = \arctan\left(\nicefrac{r_{a2}}{r_{a1}}\right)$ which means   $ \pderiv{\varphi}{r_{a1}}=\left(\frac{1}{1+(\nicefrac{r_{a2}}{r_{a1}})^2}\right)\left(\nicefrac{-r_{a2}}{r_{a1}^2}\right)$
         \mbox{$ = \left(\nicefrac{r_{a1}^2}{r_a^2}\right)\left(\nicefrac{-r_{a2}}{r_{a1}^2}\right)
	= \nicefrac{-\sin\varphi}{r_a}.$}
\\\\

\textbf{2. Second partial derivative with respect to $r_{a1}$.} 
\begin{align}
	\psderiva{f(r_a,r_b,\varphi)}{r_{a_{1}}}
        &=\bigg[\cos\varphi\pderiv{ }{r_a} - \frac{\sin \varphi}{r_a}\pderiv{ }{\varphi}\bigg]\bigg[ \cos\varphi\pderiv{f}{r_a} - \frac{\sin \varphi}{r_a}\pderiv{f}{\varphi}\bigg]\notag\\ 
        &= \cos^2\varphi\psderiva{f}{r_a}-\frac{2\sin\varphi\cos\varphi}{r_a}\psderiv{f}{r_a}{\varphi}+\frac{\sin^2\varphi}{r_a^2}\psderiva{f}{\varphi}\notag\\
        & \quad +\frac{\sin^2\varphi}{r_a}\pderiv{f}{r_a}+\frac{2\sin\varphi\cos\varphi}{r_a^2}\pderiv{f}{\varphi}\ .\\\notag
\end{align} 
\textbf{3. Second partial derivative with respect to $r_{a1}$ and $r_{b1}$.}
\begin{align}
	\psderiv{f(r_a,r_b,\varphi)}{r_{a_{1}}}{r_{b_{1}}}
 	&=\psderiv{f(r_a,r_b,\varphi)}{r_{a_{1}}}{r_b}\notag\\
	&=\cos\varphi\psderiv{f}{r_a}{r_b}-\frac{\sin\varphi}{r_a}\psderiv{f}{r_b}{\varphi}\ .\\\notag
\end{align}
\textbf{4. Partial derivative with respect to $r_{a2}$.}
\begin{align}
	\pderiv{f(r_a,r_b,\varphi)}{r_{a_{2}}}
        &= \pderiv{f}{r_a}\pderiv{r_a}{r_{a_{2}}}+ \pderiv{f}{r_b}\pderiv{r_b}{r_{a_{2}}}+ \pderiv{f}{\varphi}\pderiv{\varphi}{r_{a_{2}}}\notag\\
	 &=\sin\varphi\pderiv{f}{r_a} +\frac{\cos \varphi}{r_a}\pderiv{f}{\varphi}\ .  
\end{align} 
This uses $\pderiv{r_a}{r_{a2}} = \frac{r_{a2}}{(r_{a1}^2+ r_{a2}^2)^{\nicefrac{1}{2}}} = \sin \varphi$,    
and $\pderiv{\varphi}{r_{a2}}= \left(\frac{1}{1+(\nicefrac{r_{a2}}{r_{a1}})^2}\right)\left(\frac{1}{r_{a1}}\right)$=\mbox{ $\frac{\cos\varphi}{r_a}$}.
\\\\

\textbf{5. Second partial derivative with respect to $r_{a2}$.} 
\begin{align}
	\psderiva{f(r_a,r_b,\varphi)}{r_{a_{2}}}
          &=\bigg[\sin\varphi\pderiv{ }{r_a} + \frac{\cos\varphi}{r_a}\pderiv{ }{\varphi}\bigg]\bigg[ \sin\varphi\pderiv{f}{r_a} + \frac{\cos\varphi}{r_a}\pderiv{f}{\varphi}\bigg]\notag\\ 
	&= \sin^2\varphi\psderiva{f}{r_a}+\frac{2\sin\varphi\cos\varphi}{r_a}\psderiv{f}{r_a}{\varphi}+\frac{\cos^2\varphi}{r_a^2}\psderiva{f}{\varphi}\notag\\
        & \quad +\frac{\cos^2\varphi}{r_a}\pderiv{f}{r_a}-\frac{2\sin\varphi\cos\varphi}{r_a^2}\pderiv{f}{\varphi}\ .
 \end{align}

Having assembled these ingredients we can substitute into Eq. (\ref{eq:ABC2pd2}) to get

\begin{align}
	\langle ABC\rangle_2&=  Z_{ac}^{11}\psderiva{\langle ABC \rangle}{r_b}\notag\\
	&\quad +Z_{bc}^{11}\bigg[\cos^2\varphi\psderiva{\langle ABC \rangle}{r_a}-\frac{2\sin\varphi\cos\varphi}{r_a}\psderiv{\langle ABC \rangle}{r_a}{\varphi}+\frac{\sin^2\varphi}{r_a^2}\psderiva{\langle ABC \rangle}{\varphi}\notag\\
        & \quad +\frac{\sin^2\varphi}{r_a}\pderiv{\langle ABC \rangle}{r_a}+\frac{2\sin\varphi\cos\varphi}{r_a^2}\pderiv{\langle ABC \rangle}{\varphi}\bigg]\notag\\
	&\quad +Z_{bc}^{22}\bigg[ \sin^2\varphi\psderiva{\langle ABC \rangle}{r_a}+\frac{2\sin\varphi\cos\varphi}{r_a}\psderiv{\langle ABC \rangle}{r_a}{\varphi}+\frac{\cos^2\varphi}{r_a^2}\psderiva{\langle ABC \rangle}{\varphi}\notag\\
        & \quad +\frac{\cos^2\varphi}{r_a}\pderiv{\langle ABC \rangle}{r_a}-\frac{2\sin\varphi\cos\varphi}{r_a^2}\pderiv{\langle ABC \rangle}{\varphi}\bigg]\notag\\
	&\quad +[Z_{ac}^{11}+Z_{bc}^{11}-Z_{ab}^{11}] \bigg[\cos\varphi\psderiv{\langle ABC \rangle}{r_a}{r_b}-\frac{\sin\varphi}{r_a}\psderiv{\langle ABC \rangle}{r_b}{\varphi}\bigg]\ .\\ \notag
\end{align}
 To finish we collect together terms in each partial derivative and obtain Eq. (\ref{eq:ABC2*}), the final result for the deflection contribution to the lensed 3PCF. 
\section{The unlensed matter three-point correlation function}\label{U3PCF}
 The three-point correlation function $\zeta(\mathbf{r_1},\mathbf{r_2},\mathbf{r_3})$ of the matter density field is defined as in Eq. (\ref{eq:u3PCF}), with the matter bispectrum in place of the general bispectrum.  The matter bispectrum can be computed using Eulerian perturbation theory \cite{RN16, RN5} as 
\begin{align}
	B_{\mathrm{PT}}(\mathbf{k}_1,\mathbf{k}_2,\mathbf{k}_3)= \left[\frac{10}{7}+\frac{\mathbf{k}_1\cdot\mathbf{k}_2}{k_1k_2}\left(\frac{k_1^2+k_2^2}{k_1k_2}\right)
	+\frac{4}{7}\left(\frac{\mathbf{k}_1\cdot\mathbf{k}_2}{k_1k_2}\right)^2\right]
	P_\delta(k_1)P_\delta(k_2) + \text{2 perms.}\ ,
\end{align}
where $P_\delta(k)$ is the matter power spectrum.
\\\\
 From this it is possible to derive the following expression for the three-point correlation function \cite{RN4}:
\begin{align}
  \zeta(\mathbf{r_1},\mathbf{r_2},\mathbf{r_3})  &= \frac{10}{7} \xi(r_{21})\xi(r_{31}) - \big[\eta_2(r_{21})\eta_0(r_{31})+  \eta_0(r_{21})\eta_2(r_{31})\big]\mathbf{r}_{21}\cdot\mathbf{r}_{31}\notag\\
 	&\quad+\frac{4}{7}\big[\epsilon(r_{21})\epsilon_2(r_{31})(\mathbf{r}_{21}\cdot\mathbf{r}_{31})^2 + \epsilon(r_{21})\eta_2(r_{31})r_{21}^2\notag\\
	&\quad+\eta_2(r_{21})\epsilon(r_{31})r_{31}^2 + 3\eta_2(r_{21})\eta_2(r_{31})\big] + \text{2 perms.}
\end{align}
where $r_{ij}=|\mathbf{r}_i-\mathbf{r}_j|$. 
\\\\
The functions $\xi(r)$,  $ \eta_l(r)$ and $\epsilon(r)$ are given by 
\begin{align}
	\xi(r)&= \frac{1}{2\pi^2}\int_0^\infty \mathrm{d}k\, k^2 P_\delta(k) j_0(kr)\ ,\label{eq:xir}\\
           \eta_l(r)&= -\frac{1}{2\pi^2}\int_0^\infty \mathrm{d}k\, k^2 P_\delta(k) \frac{k}{k^lr}j_1(kr)\ ,\label{eq:etar}\\
	\epsilon(r)&= \frac{1}{2\pi^2}\int_0^\infty \mathrm{d}k\, k^2 P_\delta(k) \frac{k^2}{r^2}j_2(kr)\ .\label{eq:epsr}
\end{align}
$\xi(r)$ is the two-point correlation function. $\eta_l(r)$ has two variants with $l=0$ and $l=2$.
\\
 These expressions can be problematic to integrate numerically because the Bessel functions are oscillatory.
 However they can in fact be evaluated efficiently with Fast Fourier Transforms (FFT) \cite{RN13, Simon}.
     To achieve this we can use the fact that  it is possible to transform equations of the form 
\begin{align}
f(r) &= \int_0^\infty\mathrm{d}k\,r F(kr) \hat{f}(k)\  ,\label{eq:fr}
\end{align}
for some function $F$, with $f(r)$ and $\hat{f}(k)$ a Hankel transform pair,
\begin{align}
\hat{f}(k) =\int_0^\infty \mathrm{d}r\, k F(kr) f(r)\ ,
\end{align}
in a way that makes them readily integrable.
To see this we make a change of variables $r\equiv \mathrm{e}^x$ and $k\equiv \mathrm{e}^y$ 
and define \mbox{$g(x)\equiv f(\mathrm{e}^x)$} and \mbox{$\hat{g}(y)\equiv \hat{f}(\mathrm{e}^y)$}.
Then Eq. (\ref{eq:fr}) becomes
\begin{align}
      g(x)&=\int_{-\infty}^\infty\mathrm{d}y \; \mathrm{e}^{x+y} F(e^{x+y}) \hat{g}(y)\\
      &=\int_{-\infty}^\infty\mathrm{d}y \;  G({x+y}) \hat{g}(y)\ ,\label{eq:cc}
\end{align}
where ${G}(z)$ is defined as $\mathrm{e}^zF(\mathrm{e}^z)$.
\\\\
The integral in Eq. (\ref{eq:cc}) is the cross-correlation $G\star \hat{g}(y)$ which is equivalent to the convolution $G\ast\hat{ g}^\ast(-y)$ where $ \hat{g}^\ast$ is the complex conjugate of $\hat{g}$. We can thus avoid the need for integration by  transforming to Fourier space where convolutions become products.

To transform Eqs. (\ref{eq:xir}-\ref{eq:epsr}) to the required form we change the spherical Bessel functions to cylindrical Bessel functions using the identity
\begin{align}
 	j_\mu(z)=\sqrt{\frac{\pi}{2z}}J_{\mu+\frac{1}{2}}(z)\ .
\end{align}

We can then write the equations as:
\begin{align}
	\xi(r)&= \sqrt{\frac{\pi}{2}}\frac{1}{2\pi^2}r^{-1}\int_0^\infty  \mathrm{d}k\; k^2 P_\delta(k) r(kr)^{-\frac{1}{2}}J_{\frac{1}{2}}(kr)\ ,\label{eq:xi}\\
	\eta_0(r)&= -\sqrt{\frac{\pi}{2}}\frac{1}{2\pi^2}r^{-3}\int_0^\infty  \mathrm{d}k\; k^2 P_\delta(k)r(kr)^{\frac{1}{2}}J_{\frac{3}{2}}(kr)\ ,\\
	\eta_2(r)&= -\sqrt{\frac{\pi}{2}}\frac{1}{2\pi^2}r^{-1}\int_0^\infty  \mathrm{d}k\; k^2 P_\delta(k)r(kr)^{-\frac{3}{2}}J_{\frac{3}{2}}(kr)\ ,\\
	\epsilon(r)&= \sqrt{\frac{\pi}{2}}\frac{1}{2\pi^2}r^{-3}\int_0^\infty  \mathrm{d}k\; k^2 P_\delta(k)r(kr)^{\frac{1}{2}}J_{\frac{5}{2}}(kr)\ .\label{eq:eps2}
\end{align}
These have the form of Eq. (\ref{eq:fr}). 

\end{document}